\documentclass[
    aps,
    pra,
    reprint,
    amsmath,
    amssymb,
    longbibliography
]{revtex4-2}

\usepackage{graphicx}
\usepackage{dcolumn}
\usepackage{bm}

\usepackage{amsmath}
\usepackage{old-arrows}
\usepackage{comment}
\usepackage{multirow}
\usepackage{tikz-cd}
\usepackage{enumitem}
\usepackage{stmaryrd}
\usepackage{braket}
\usepackage{url}
\usepackage{amsthm}
\usepackage[ruled,vlined, linesnumbered]{algorithm2e} 
\SetKwInput{KwRequire}{Require}

\newcommand{\Z}{\mathbb{Z}}

\newcommand{\R}{\mathbb{R}}
\newcommand{\C}{\mathbb{C}}

\newcommand{\calC}{\mathcal{C}}
\newcommand{\calH}{\mathcal{H}}
\newcommand{\calB}{\mathcal{B}}
\newcommand{\eg}{{\em e.g.}}
\newcommand{\ie}{{\em i.e.}}

\DeclareMathOperator{\SU}{SU}

\DeclareMathOperator{\Ad}{Ad}

\DeclareMathOperator{\Span}{span}

\DeclareMathOperator{\tr}{tr}

\DeclareMathOperator{\Lie}{Lie}
\DeclareMathOperator{\diag}{diag}
\DeclareMathOperator{\sinc}{sinc}
\DeclareMathOperator{\su}{\mathfrak{su}}

\newcommand{\calP}{\mathcal{P}}
\newcommand{\calM}{\mathcal{M}}

\newtheorem{theorem}[equation]{Theorem}

\newtheorem{lemma}[equation]{Lemma}

\newtheorem{proposition}[equation]{Proposition}
\theoremstyle{remark}
\newtheorem{remark}[equation]{Remark}

\begin{document}

\preprint{APS/123-QED}

\title{\textbf{Theory of Trotter Error with Variable-Weight Schedules} 
}%

\author{Henrique Ennes}
 \email{Contact author: henrique.lovisi-ennes@inria.fr}
\affiliation{Universit\'e C\^ote d'Azur, Inria, CNRS, I3S, Sophia Antipolis, France
}%

\date{\today}

\begin{abstract}
Trotterization is a widely used algorithm for simulating quantum evolutions and consists of approximating a target unitary through a local linear decomposition of its generator that is repeated over $N$ steps.
In physical implementations, the coefficients of this decomposition may correspond to experimentally tunable quantities, such as pulse durations or amplitudes.
We introduce scheduled trotterization, a variable-weight modification that exploits this control freedom by allowing the coefficients to change from step to step while preserving the underlying pulse ordering.
Using the geometry of the Lie group of unitary operators, we express the global endpoint error as a product of transported local defects, whose leading terms are obtained by two applications of the Baker-Campbell-Hausdorff formula, one inside each local step and one between different transported defects, separating the internal Trotter error from the external accumulation error.
While we recover recent tighter bounds on the global approximation error when the weights are held fixed, the new framework allows us to choose a weight-varying schedule that coherently suppresses the accumulated error.
For exact infinitesimal synthesis, we prove that this improves the usual fixed-weight $O(N^{-1})$ behavior to $O(N^{-2})$, without changing the basic pulse skeleton.
We also extend the framework to imperfect infinitesimal synthesis, as arises in quantum gate synthesis problems, where admissible coefficients may only approximate the desired local corrections.
In this case, the error is controlled by a local covering parameter measuring the quality of the available coefficient set.
Numerical experiments for random $\SU(2)$ targets further support the predicted theoretical improvements.
The results suggest a novel perspective for the analysis of Trotter error, in which experimentally available control freedom can be used to coherently cancel its accumulation at the endpoint.
\end{abstract}

\maketitle

\section{Introduction}
\emph{Product formulas} correspond to a family of algorithms for decomposing quantum dynamics using a restricted set of experimentally available controls.
A target evolution generated by a Hamiltonian $H$ is decomposed into a sequence of simpler evolutions called \emph{pulses}, which can be directly implemented in the laboratory and whose corresponding Hamiltonians, $H_1,\dots,H_M$, decompose $H$ as a linear combination.
Depending on the physical platform, the \emph{weights} of these linear combinations have different physical implementations, such as pulse durations, amplitudes, or repeated applications of elementary quantum gates.
Because of their simplicity and low control overhead, product formulas have become standard tools in quantum simulation~\cite{feynman2018simulating, whitfield2011simulation,wecker2014gate,hastings2015improving, poulin2015trotter,babbush2015chemical,kivlichan2020improved, motta2021low,su2021nearly, smith2019simulating, haah2018quantum, doi:10.1126/science.273.5278.1073, childs2018toward}, quantum control~\cite{d2021introduction, khaneja2001time,nielsen2002universal, childs2013product,somma2016trotter, weimer2010rydberg}, gate synthesis~\cite{deutsch1995universality, doi:10.1126/science.1208001,ennes2026removingonlineexponentialnet}, Monte Carlo problems~\cite{de1983applications, schmidt1995high, JANKE1992199}, and even in numerical analysis~\cite{blanes2024splitting, hairer2006geometric}, where they are known as splitting methods.

\begin{figure*}[t]
\centering
\includegraphics[width=0.9\textwidth]{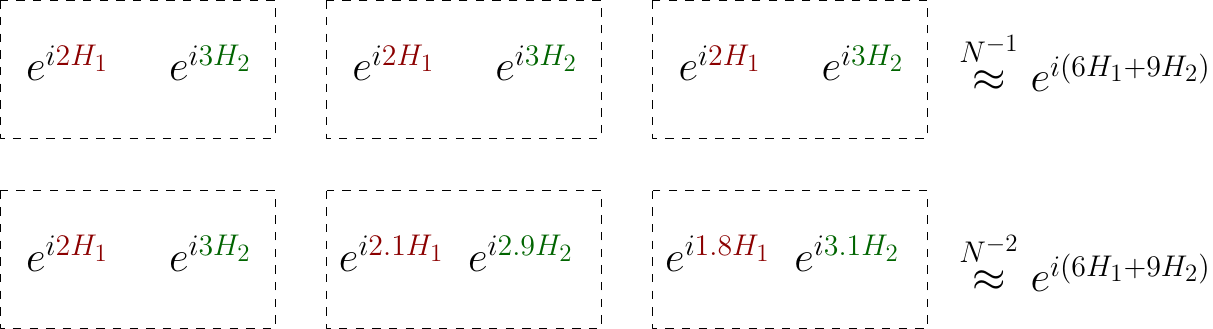}
\caption{\textbf{Top}: Ordinary fixed-weight trotterization of the unitary $U=\exp[i(6H_1+9H_2)]$ using the pulses $\exp(i2H_1)$ and $\exp(i3H_2)$. The different steps are separated by the dashed rectangles; here the number of steps $N$ is set to 3 for illustration purposes. \textbf{Bottom}: Variable-weight trotterization of the same target unitary, where the pulse weights are adjusted at later steps to suppress the accumulated error. While the fixed-weight trotterization solution converges to the target evolution with scaling $O(N^{-1})$, our variable-weight schedule converges with scaling $O(N^{-2})$.}
\label{fig: skeleton}
\end{figure*}

Historically, the first of the product formulas---albeit still amply used in practical applications~\cite{kim2023evidence,yoshioka2025krylov,fuller2026operator}---is what physicists know as \emph{trotterization}.
In simple terms, it rescales $H$ by a $1/N$ factor, where the positive integer $N$ controls the \emph{number of steps} of the scheme.
Therefore, instead of directly approximating the target Hamiltonian with the pulse generators, it decomposes the smaller (and hence better behaved) $H/N$, and the full evolution can be achieved by concatenating $N$ of these steps, as illustrated in Figure~\ref{fig: skeleton}.
For fixed target and pulses, the approximation error of the scheme scales as $O(1/N)$, although, in practice, this is known to be a rather loose bound in the more general case.
Higher-order product formulas~\cite{hatano2005finding, malevzivc2026efficient, morales2022greatly, omelyan2002optimized, ostmeyer2023optimised, suzuki1976generalized, yoshida1990construction}, (randomized) permutations of non-commuting Hamiltonians~\cite{anastasiou2024tetris, PhysRevLett.123.070503, chen2021concentration, chen2022efficient, childs2019faster, decker2026kernpiler, li2022paulihedral, poulin2011quantum, yan2026physics, zhang2012randomized}, or multiproduct extrapolation methods~\cite{aftab2024multi, carreraVazquez2023wellconditioned, faehrmann2022randomizing, low2019well, zhuk2024trotter} can improve this scaling, but this generally requires more quantum resources or more complicated control skeletons.

There is, however, another degree of freedom that is often available experimentally.
Even when the ordering of the control Hamiltonians is kept fixed, their durations or amplitudes need not be identical at each Trotter step.
This suggests a natural question: can these already available control degrees of freedom be used to suppress Trotter error without modifying the underlying pulse sequence?

In this work, we show that they can.
To this end, we consider \emph{scheduled trotterization}, in which the coefficients of the local product formula are allowed to vary from step to step while the pulse ordering is preserved.
The key observation is that local Trotter errors do not simply add up in norm; they are coherently transported along the ideal evolution.
By tracking this transport, the weights of a new step can be chosen so that its \emph{controlled drift} compensates for the leading error accumulated during the previous steps.
In particular, we show that when the available control Hamiltonians span a Lie algebra and the required infinitesimal corrections can be synthesized exactly, the resulting schedule improves the global first-order Trotter error from $O(N^{-1})$ to $O(N^{-2})$.
In other words, the resulting accuracy has the same asymptotic dependence on $N$ as second-order symmetric product formulas, but is achieved by slightly modulating the weights of the original first-order pulse skeleton rather than by replacing it with a higher-order composition.

We also consider more general settings in which the admissible controls are restricted.
This includes gate synthesis, where the coefficients are integer-valued because they count repeated applications of elementary gates and their inverses.
We introduce a local covering parameter that quantifies the resulting imperfect infinitesimal synthesis and derive corresponding global error bounds.
Numerical experiments on $\SU(2)$ illustrate both the quadratic improvement in the continuously controlled regime and the behavior of the method under discrete gate constraints.
Moreover, when the available Hamiltonians do not themselves span a Lie algebra, we show how the control set can be enlarged through a preprocessing procedure based on adjoint conjugations.
This extends the scheduling framework to general controllable systems at the price of modifying the pulse skeleton.

We should note that other methods have also considered adaptive versions of trotterization.
In~\cite{zhao2023making}, for example, the trotterization step size is adapted at each time step in order to reduce the accumulated error.
This can be viewed as a special case of our variable-weight framework, where, instead of individually adjusting each coefficient, they simultaneously rescales all weights by the same value while enforcing physical constraints such as energy conservation and gauge symmetries.
Similarly, geometric approaches have been used to design high-fidelity product formulas by optimizing the order of elementary operations; in particular, \cite{liu2020high} represents product formulas as paths and relates low-order BCH errors to geometric quantities such as areas and moments.
Other works use state information to learn adaptive product formulas~\cite{zhang2023low}, introduce spurious symmetry gates to bound error accumulation~\cite{tran2021faster}, design input states less prone to errors~\cite{zhou2026trotter}, or use variational methods for choosing ideal coefficients~\cite{assi2026beyond}.
Nevertheless, as far as we can tell, the algorithmic use of drift induced by varying weights for error cancellation is new.

The paper starts in Section~\ref{sec: variable weight} with an estimate of the trotterization error for variable-weight schedules, formalized in Theorem~\ref{th: variable trotter error}.
We then develop, in Section~\ref{sec: algorithm}, an algorithm for computing an ideal schedule.
This algorithm is quite general, but assumes that the span of the pulse generators is closed under commutators.
In Section~\ref{sec: non lie closure}, we extend the construction to the case where this closure assumption fails, at the price of modifying the pulse skeleton.
We close with a discussion of broader perspectives, followed by appendices containing the necessary Lie-theoretic background and omitted technical details.

Before closing the introduction, we stress two limitations of the present work.
First, although we conjecture that similar scheduling ideas may apply to more general product formulas, we restrict the analysis here to first-order trotterization due to its simplicity.
Second, while the geometric formalism of Section~\ref{sec: variable weight} applies to fairly general pulse sets and targets, the scheduling algorithm studied here assumes that the relevant Hamiltonians and transported residuals can be represented explicitly in a finite-dimensional basis without exploiting sparsity.
This is natural for dense control problems and low-dimensional unitary synthesis, but it does not directly give an efficient sparse-Pauli simulation algorithm.
We return to the question of variable-weight simulation of sparse operators in the conclusion.

\section{Simulating Hamiltonian evolution with variable weights}\label{sec: variable weight}
\subsection{Mathematical setup}
Consider a laboratory setting in which we have access to a finite number of {available Hamiltonians},
\begin{equation*}
    \calH=\{H_1,\dots,H_M\},
\end{equation*}
whose corresponding \emph{pulses} are
\begin{equation}\label{eq: pulse}
    \calP:=\{\exp(icH_j)\mid 1\leq j\leq M,\ c\in\calC_j\}.
\end{equation}
The set $\calH$ is a finite subset of $\su(d)$, the Lie algebra of Hermitian, traceless $d\times d$ complex matrices under our convention (refer to Appendix~\ref{sec: lie theory} for a review of the Lie-theoretic conventions used throughout the paper).
We call the elements of $\calH$ the system's \emph{available Hamiltonians}, whereas $\calC_j\subseteq\R$ is the $j$-th set of \emph{admissible weights} (or \emph{coefficients}).

As mentioned in the introduction, different laboratory settings are modeled by changing either $\calH$ or the sets $\calC_j$.
For example, the application of a constant control Hamiltonian $H$ for a time interval $t$ can be represented by the unitary operator $\exp(itH)$, in which case the admissible coefficients describe the available pulse durations or amplitudes (due to compactness of $\SU(d)$, negative-time evolutions can be approximated arbitrarily well by sufficiently long positive-time evolutions).
On the other hand, $n$ sequential applications of the same quantum circuit gate $G$ can always be written as $G^n=\exp(inH)$, where $H$ is some Hermitian matrix such that $G=\exp(iH)$; in this case, integer coefficients describe repeated applications of the gate or its inverse.
For simplicity, we denote the full set of admissible weights by
\begin{equation*}
    \calC:=\calC_1\times\dots\times\calC_M.
\end{equation*}

Suppose we are given a \emph{target unitary} $U=\exp(itH)$ which we wish to physically implement using the available set of pulses.
For simplicity, however, we fix the total evolution time to $t=1$ throughout.
We will also assume that the \emph{target Hamiltonian} $H$ belongs to the real span of $\calH$, $\Span_{\R}\calH$, although we note that the extensions considered in Section~\ref{sec: non lie closure} can be used to slightly relax this assumption.
In particular, this implies that there exists a vector $\widetilde{\alpha}\in\R^M$ such that
\begin{equation*}
    H=\sum_{j=1}^M\widetilde{\alpha}_jH_j.
\end{equation*}

The \emph{product formula synthesis problem} consists of finding a unitary
\begin{equation*}
\begin{split}
    W:=\exp(ic_{j_T}H_{j_T})\dots\exp(ic_{j_1}H_{j_1}),
\end{split}
\end{equation*}
for $H_{j_t}\in\calH,
    c_{j_t}\in\calC_{j_t},\text{ and }
    1\leq t\leq T,$ 
such that the \emph{global approximation error}
\begin{equation*}
    \varepsilon:=\|U-W\|
\end{equation*}
is small, where the norm is assumed to be the operator norm.
Note that $W$ can be fully described by the order in which each available Hamiltonian is applied, $H_{j_1},\dots,H_{j_T}$, and the corresponding list of weights, $c_{j_1},\dots,c_{j_T}$.
The former is called $W$'s \emph{skeleton} and the latter its corresponding \emph{schedule}.

Trotterization provides an easy-to-compute solution to the product formula synthesis problem.
Choose a large positive integer $N$, whose size controls the approximation accuracy.
For simplicity, we write
\begin{equation}\label{eq: hamiltonian decomposition}
    \frac{H}{N}:=\sum_{j=1}^M\alpha_jH_j,
\end{equation}
that is, $\alpha_j:=\widetilde{\alpha}_j/N$.
We then let the full \emph{trotterization} be
\begin{equation*}
    W:=
    \overleftarrow{\prod_{n=1}^N}
    \left(
    \overrightarrow{\prod_{j=1}^M}
    \exp(ic_j^{(n)}H_j)
    \right),
    \qquad
    c_j^{(n)}\in\calC_j
\end{equation*}
for all $1\leq j\leq M$ and $1\leq n\leq N$, where the arrows indicate the order of the products.
Explicitly,
\begin{equation*}
\begin{split}
     W =&
    \exp(ic_1^{(N)}H_1)\dots\exp(ic_M^{(N)}H_M)\\
    & \;\;\dots\exp(ic_1^{(1)}H_1)\dots\exp(ic_M^{(1)}H_M). 
\end{split}
\end{equation*}

\begin{figure}
\centering
\includegraphics[width=0.45\textwidth]{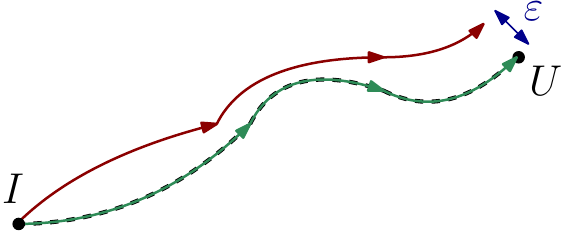}
\caption{A representation of the geometric intuition behind trotterization. The black dashed curve corresponds to the path $\gamma(t)=\exp(itH)$ in $\SU(d)$, starting at the identity when $t=0$ and ending at $\gamma(1)=U$. The ideal integration curve (green), $S^n=\exp(inH/N)$, coincides with $\gamma(n/N)$ and therefore incurs no approximation error when $n=N$. On the other hand, a general trotterization scheme (red) uses actual steps $T^{(n)}$ and, at $n=N$, ends at distance $\varepsilon$ (blue) from the target. Here $N=3$ for illustration purposes.}
\label{fig: paths}
\end{figure}

Note that there is some freedom in choosing the order of the exponentials within each step, but this order should be the same throughout all $N$ steps.
For later reference, we denote each such step by the unitary
\begin{equation*}
    T^{(n)}
    :=
    \overrightarrow{\prod_{j=1}^M}\exp(ic_j^{(n)}H_j),
\end{equation*}
which we call the \emph{actual} trotterization step.
The word ``step" refers here to the fact that trotterization works similarly to the numerical integration of a path in $\SU(d)$.
Just as, say, in Euler's method, a path is described by a sequence of consecutive steps, $W$ can be seen as the sequential application of the steps $T^{(n)}$.
``Actual", on the other hand, is used to distinguish $T^{(n)}$ from the \emph{ideal step},
\begin{equation*}
    S:=\exp(iH/N),
\end{equation*}
for which $S^N=U$, see Figure~\ref{fig: paths}.

Clearly, if $S=T^{(n)}$ for all $n$, then trotterization exactly synthesizes $U$, but, in practice, exactly implementing $S$ with the laboratory apparatus is usually impossible.
Still, we want to make $T^{(n)}$ as close as possible to $S$ and define the trotterization \emph{local discrepancy} as
\begin{equation}\label{eq: discrepancy}
    D^{(n)}:=S^{-1}T^{(n)}.
\end{equation}
We also define the \emph{infinitesimal discrepancy} as its logarithm,
\begin{equation*}
    X^{(n)}:=\log(D^{(n)})\in\su(d).
\end{equation*}
Finally, we let the $n$-th \emph{drift} of a schedule
$\{c_j^{(n)}\}_{1\leq j\leq M,\ 1\leq n\leq N}$ be
\begin{equation*}
    \Delta^{(n)}
    :=
    \sum_{j=1}^M(c_j^{(n)}-\alpha_j)H_j.
\end{equation*}
We will relate the infinitesimal discrepancy and the drift in the next section.

\subsection{Estimating variable-weight trotterization error}

For a given schedule
$\{c_j^{(n)}\}_{1\leq j\leq M,\ 1\leq n\leq N}$, the error accumulated up to step $n$ is
\begin{equation}\label{eq: epsilon n}
    \varepsilon^{(n)}
    :=
    \left\|
    S^n-\overleftarrow{\prod_{t=1}^n}T^{(t)}
    \right\|
    =
    \left\|
    S^{-n}\overleftarrow{\prod_{t=1}^n}T^{(t)}-I
    \right\|.
\end{equation}
In particular, the total accumulated error for this schedule is simply
$\varepsilon=\varepsilon^{(N)}$.
We can estimate this quantity by noting that
\begin{equation*}
\begin{split}
S^{-n}\overleftarrow{\prod_{t=1}^n}T^{(t)}
&=
S^{-n}SD^{(n)}SD^{(n-1)}\dots SD^{(1)}\\
&=\overleftarrow{\prod_{t=1}^n}
S^{1-t}D^{(t)}S^{t-1}\\
&=
\exp\left(i\Ad_{S^{1-n}}X^{(n)}\right)
\dots
\exp\left(iX^{(1)}\right).
\end{split}
\end{equation*}
Here we used equation~\eqref{eq: discrepancy}, followed by the definition of the
\emph{adjoint action} (see Appendix~\ref{sec: lie theory}).
Equivalently, this can be written more compactly as
\begin{equation}\label{eq: transport identity}
S^{-n}\overleftarrow{\prod_{t=1}^n}T^{(t)}
=
\overleftarrow{\prod_{t=1}^n}
\exp\left[i\Ad_{S^{1-t}}X^{(t)}\right].
\end{equation}

Geometrically, equation~\eqref{eq: transport identity} shows that the accumulated failure to approximate the ideal path up to step $n$ is the composition of the local defects, with their infinitesimal forms transported to a common frame by the adjoint action of $\SU(d)$ on $\su(d)$.
For this reason, we shall refer to it as the \emph{transport identity}.
In fact, disguised versions of this relation for fixed weights were already known in the literature; see, for example,~\cite{chen2025trotter, tran2020destructive}.
Applying the transport identity to equation~\eqref{eq: epsilon n} and using Lemma~\ref{lm: properties operator norm}, we bound the total error by
\begin{equation}\label{eq: log bound}
    \varepsilon
    \leq
    \left\|
    \log\left(
    \overleftarrow{\prod_{n=1}^N}
    \exp\left(i\Ad_{S^{1-n}}X^{(n)}\right)
    \right)
    \right\|.
\end{equation}

Analyzing equation~\eqref{eq: log bound}, we encounter a nested pair of matrix logarithms.
The first is an \emph{external} logarithm acting on the product of exponentials of $X^{(n)}$ (modulo the adjoint action), while the second is an \emph{internal} logarithm, recalling that
$X^{(n)}=\log(S^{-1}T^{(n)})$.
Provided that the corresponding BCH expansions are sufficiently close to the identity of $\SU(d)$, the \emph{Baker-Campbell-Hausdorff} (\emph{BCH}) formula~\cite{hall2013lie} can be used to transform the products involved into sums of Hermitian operators.

Explicitly, if $Y_1,\dots,Y_M$ are Hermitian operators satisfying
$$
    \sum_{i=1}^M\|Y_i\|\leq\delta,
$$
for some $\delta\leq\delta_0$, where $0<\delta_0<1$ is a sufficiently small universal constant, then
\begin{equation}\label{eq: bch 1}
    \log\left[\overrightarrow{\prod_{i=1}^M}\exp(iY_i) \right] =
    \sum_{i=1}^M Y_i + \frac{1}{2}\sum_{i<j}[Y_i,Y_j] + R_3,
\end{equation}
where the remainder $R_3\in\su(d)$ satisfies
$$
    \|R_3\|\leq C\delta^3,
$$
for some universal constant $C$, independent of $d$ and $M$.
Here, we use the Hermitian convention for the Lie bracket,
\begin{equation*}
    [Y_i,Y_j]:=i(Y_iY_j-Y_jY_i).
\end{equation*}
The assumption $\delta_0<1$ is made for later convenience; for the BCH expansion itself, one may take $\delta_0<\ln 2$~\cite[Proposition 2.2]{biagi2020baker}.

To apply BCH to the local discrepancy, we require
\begin{equation*}
    \left\|\frac{H}{N}\right\| + \sum_{j=1}^M\|c_j^{(n)}H_j\| \leq\delta_0,
\end{equation*}
while the external BCH expansion requires
\begin{equation*}
    \sum_{n=1}^N\|X^{(n)}\|\leq\delta_0.
\end{equation*}
The former of these conditions can be enforced by letting $N$ large, whereas the second will be true provided that the vectors $c^{(n)}$ and $\alpha$ are close to each other---\ie, that the drift is small.
When they hold, the inner logarithm becomes
\begin{equation*}
\begin{aligned}
\log(S^{-1}T^{(n)})
&=
\log\left(
\exp\left(-i\frac{H}{N}\right)
\overrightarrow{\prod_{j=1}^M}
\exp\left(ic_j^{(n)}H_j\right)
\right)
\\
&\approx
\Delta^{(n)}
+
\frac{1}{2}\sum_{i<j}
\left[c_i^{(n)}H_i,c_j^{(n)}H_j\right]
\\
&\qquad
-\frac{1}{2N}[H,\Delta^{(n)}].
\end{aligned}
\end{equation*}
where we used
$\sum_jc_j^{(n)}H_j=H/N+\Delta^{(n)}$.
Similarly, the outer approximates to
\begin{equation*}
\begin{aligned}
\sum_{n=1}^N
\Ad_{S^{1-n}}X^{(n)}
+
\frac{1}{2}\sum_{m<n}
\left[
\Ad_{S^{1-n}}X^{(n)},
\Ad_{S^{1-m}}X^{(m)}
\right]
\end{aligned}
\end{equation*}
up to higher-order terms.
We formalize these estimates in Theorem~\ref{th: variable trotter error}, whose proof can be found in Appendix~\ref{sec: proof error bound}.

\begin{theorem}[Scheduled trotterization error]\label{th: variable trotter error}
Let $c^{(1)},\dots,c^{(N)}\in\calC$ be a weight schedule for the target
$U=e^{iH}$, and suppose that, for every $1\leq n\leq N$,
\begin{equation*}
    \left\|\frac{H}{N}\right\|\leq\delta, \qquad \sum_{j=1}^M\|c_j^{(n)}H_j\|\leq\delta,
\end{equation*}
with $2\delta\leq\delta_0$.
Define
\begin{equation*}
    \sigma:=\sum_{n=1}^N\|X^{(n)}\|, \qquad \beta:=\max_{1\leq n\leq N}\|\Delta^{(n)}\|,
\end{equation*}
and assume $\sigma\leq\delta_0$.
Then
\begin{equation}\label{eq: full global error}
    \varepsilon \leq \|R_{\text{int}}\| + \|R_{\text{ext}}\| + C\left( N\delta^3 + N^2\delta^2\beta + N^2\delta^4 + \sigma^3 \right),
\end{equation}
where
\begin{equation}\label{eq: remainder terms}
\begin{aligned}
R_{\text{int}}
&=
\sum_{n=1}^N
\Ad_{S^{1-n}}
\Bigg(
\Delta^{(n)}
+
\frac{1}{2}\sum_{i<j}
[c_i^{(n)}H_i,c_j^{(n)}H_j]
\\
&\qquad\qquad
-\frac{1}{2N}[H,\Delta^{(n)}]
\Bigg),
\\[0.5em]
R_{\text{ext}}
&=
\frac{1}{2}\sum_{m<n}
\left[
\Ad_{S^{1-n}}\Delta^{(n)},
\Ad_{S^{1-m}}\Delta^{(m)}
\right].
\end{aligned}
\end{equation}
Here $C$ and $\delta_0$ are universal constants.
\end{theorem}

\subsection{Fixed-weight solution}

Theorem~\ref{th: variable trotter error} applies to variable-weight schedules, but it is instructive to specialize it to the case where $c^{(n)}=c$ is a fixed vector in $\R^M$, as this corresponds exactly to the usual trotterization scheme.
Even in the fixed-weight setting, directly minimizing equation~\eqref{eq: full global error} does not seem trivial, so it is usual to choose $c$ to minimize the drift, that is,
\begin{equation}\label{eq: c as min}
    c\in \arg\min_{c\in\calC}\|\Delta\|.
\end{equation}
Whenever this quantity can be set exactly to zero---that is, whenever $\alpha\in\calC$---we say that the fixed-weight solution is \emph{drift-free}.

Plugging this into equations~\eqref{eq: full global error} and~\eqref{eq: remainder terms}, and using the usual fixed-weight scaling $\delta=O(N^{-1})$, we get
\begin{equation}\label{eq: error drift free}
\begin{split}
      \varepsilon_{\text{fixed-weight},\Delta=0} &\leq \frac{1}{2} \left\| \sum_{n=1}^N\sum_{i<j} \Ad_{S^{1-n}} \left[\alpha_iH_i,\alpha_jH_j\right] \right\| \\
      & \quad + O(N\delta^3).
\end{split}
\end{equation}
Under the crude bound obtained by applying the triangle inequality, equation~\eqref{eq: error drift free} agrees, to leading order, with the traditional trotterization error estimate in the literature,
\begin{equation}\label{eq: crude bound}
\begin{split}
    \varepsilon_{\text{fixed-weight},\Delta=0,\text{crude}} \lesssim \frac{N}{2} \left\| \sum_{i<j} \left[\alpha_iH_i,\alpha_jH_j\right]\right\|.
\end{split}
\end{equation}
In general, however, equation~\eqref{eq: error drift free} may give a much finer estimate, as inter-step errors can coherently add up and cancel.
This behavior can be traced back to our geometric use of $\SU(d)$: while the usual argument~\cite{ennes2026removingonlineexponentialnet} makes direct use of the group's embedding in the Euclidean space of $d\times d$ complex matrices, our approach uses the adjoint action of $\SU(d)$ on $\su(d)$ to transport the local errors along the path defined by the ideal trotterization to a common frame, where the addition of errors at distinct steps is well-defined.
Again, some form of coherent interaction of the accumulated trotterization error has been repeatedly suggested in the literature~\cite{chen2026general, chen2025trotter, malevzivc2026efficient, maxwell2026practical, tran2020destructive, zhou2026trotter} to explain the observed discrepancy between empirical errors and theoretically predicted error bounds such as equation~\eqref{eq: crude bound}~\cite{childs2018toward, PhysRevX.11.011020, zhao2022hamiltonian}.
In fact, equation~\eqref{eq: error drift free} was previously derived as Eqs.~(3) and (B5) in~\cite{chen2025trotter} using a telescoping argument and as Eq.~(6) in~\cite{zhou2026trotter} through perturbative techniques.

We can obtain more refined estimates of equation~\eqref{eq: error drift free} by introducing a coordinate system on $\C^d$.
In particular, since $H$ is Hermitian, the spectral theorem guarantees the existence of a basis of $\C^d$ in which $H$ is diagonal, with entries given by its eigenvalues $\lambda_1,\dots,\lambda_d$, so that
\begin{equation*}
    S=\diag(e^{i\lambda_1/N},\dots,e^{i\lambda_d/N}).
\end{equation*}
If we denote by $\left[\alpha_iH_i,\alpha_jH_j\right]_{ab}$ the $ab$-entry of the matrix $\left[\alpha_iH_i,\alpha_jH_j\right]$ in the $H$-eigenbasis, then
\begin{equation*}
\begin{aligned}
\left(
\Ad_{S^{1-n}}
[\alpha_iH_i,\alpha_jH_j]
\right)_{ab}
&=
\left(
S^{1-n}
[\alpha_iH_i,\alpha_jH_j]
S^{n-1}
\right)_{ab}
\\
&=
e^{i(1-n)\lambda_a/N}
[\alpha_iH_i,\alpha_jH_j]_{ab}
\\
&\qquad\times
e^{-i(1-n)\lambda_b/N}
\\
&=
e^{i(1-n)\omega_{ab}/N}
[\alpha_iH_i,\alpha_jH_j]_{ab}.
\end{aligned}
\end{equation*}
where $\omega_{ab}:=\lambda_a-\lambda_b$.
Plugging this into equation~\eqref{eq: error drift free} and using the geometric sum formula gives
\begin{equation}\label{eq: fourier}
    \varepsilon_{\text{fixed-weight},\Delta=0} \lesssim \frac{1}{2}\|K\| + O(N\delta^3),
\end{equation}
where $K$ is the matrix whose entries in the $H$-eigenbasis are
\begin{equation*}
    (K)_{ab} := e^{-i\frac{N-1}{2N}\omega_{ab}} \frac{\sin(\omega_{ab}/2)}{\sin(\omega_{ab}/2N)} \sum_{i<j} \left[\alpha_iH_i,\alpha_jH_j\right]_{ab}.
\end{equation*}
In other words, the eigengaps $\omega_{ab}$ determine the degree of cancellation of each $ab$-entry of the commutator contribution under geometric transport.
For example, when $\lambda_a=\lambda_b$ (\ie, the corresponding eigenvalues are degenerate), the geometric sum is $N$ and the leading error reduces exactly to its crude estimate.
For large $N$ and fixed $\omega_{ab}$,
\begin{equation*}
    \frac{\sin(\omega_{ab}/2)}{\sin(\omega_{ab}/2N)} = N\sinc(\omega_{ab}/2)+O(N^{-1}),
\end{equation*}
and therefore, recalling that $\alpha_j=\widetilde{\alpha}_j/N$,
\begin{equation*}
    \varepsilon_{\text{fixed-weight},\Delta=0} \lesssim \frac{1}{2N}\|\widetilde K\| + O(N^{-2}),
\end{equation*}
where
\begin{equation*}
    (\widetilde K)_{ab} := e^{-i\omega_{ab}/2} \sinc(\omega_{ab}/2) \sum_{i<j} \left[\widetilde{\alpha}_iH_i,\widetilde{\alpha}_jH_j\right]_{ab},
\end{equation*}
and $\sinc(x)=\sin(x)/x$~\cite{olver2010nist}.

\begin{remark}
    Pushing further the connection to signal processing, we may notice that the normalized geometric factor
    \begin{equation*}
        \frac{1}{N}\sum_{n=1}^N e^{i(1-n)\omega_{ab}/N}
    \end{equation*}
    can be interpreted as a finite-sample Fourier transform of a rectangular window.
    As $N\to\infty$, it converges to
    \begin{equation*}
        \int_0^1e^{-it\omega_{ab}}\,dt = e^{-i\omega_{ab}/2}\sinc(\omega_{ab}/2),
    \end{equation*}
    explaining the appearance of the sinc function~\cite{owen2007practical}.
    In this sense, geometric transport acts as a \emph{filter} on the fixed-weight commutator coefficients.
\end{remark}

\begin{figure}[t]
    \centering
    \includegraphics[width=0.48\textwidth]{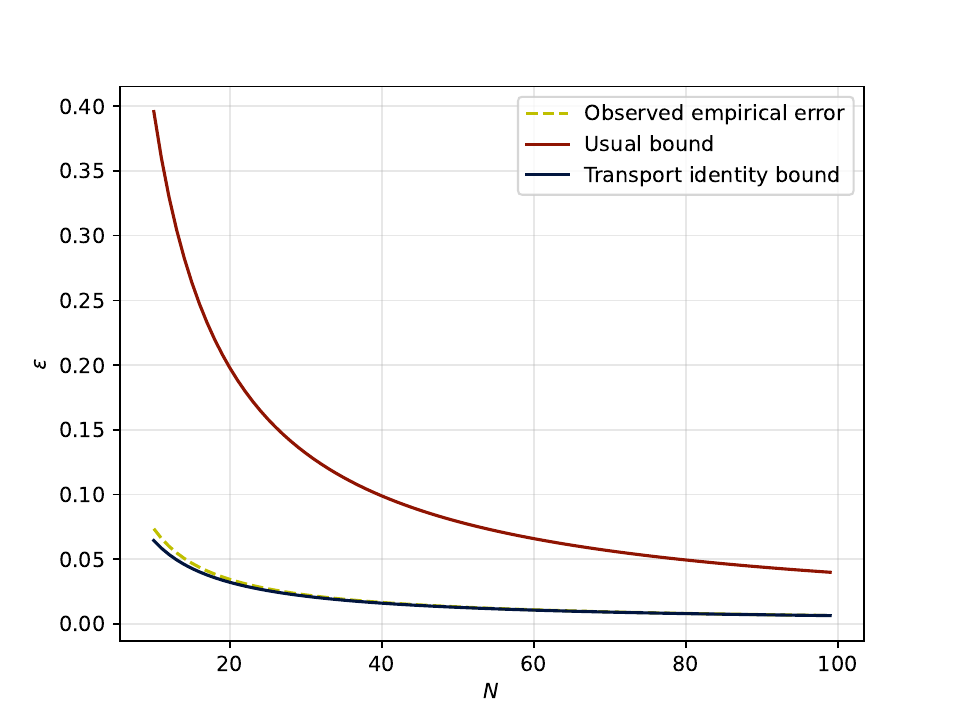}
    \hfill
    \includegraphics[width=0.48\textwidth]{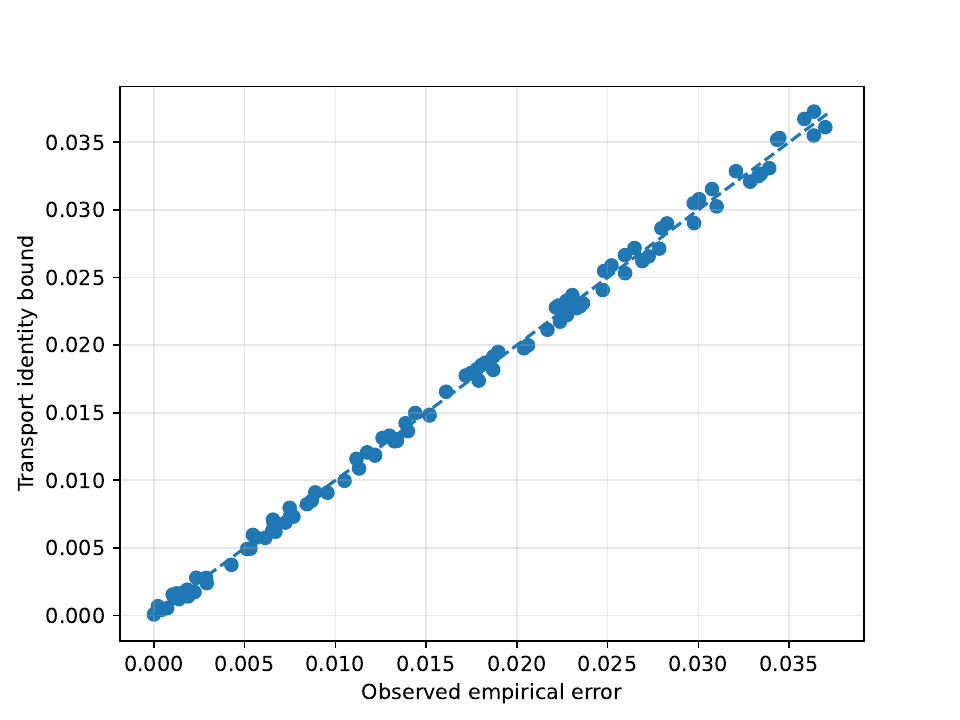}
    \caption{\textbf{Left:} Estimated and observed errors according to different trotterization bounds for different values of $N$. Here, a random $\SU(2)$ target is approximated with a fixed-weight trotterization solution, where $\calH$ is assumed to be the set of three Pauli matrices and $\calC=\R^3$ (\ie, the solution is drift-free). \textbf{Right:} The error bound estimated according to equation~\eqref{eq: fourier} as a function of the observed error, for 100 randomly sampled $\SU(2)$ targets, with $N$ held fixed at 50. The dashed line has slope $+1$.}
    \label{fig: fixed weights}
\end{figure}

Numerical experiments~\footnote{The code is publicly available at \protect\url{https://github.com/HLovisiEnnes/VarTrotter}.} suggest that equation~\eqref{eq: fourier} not only gives a tighter bound, but also accounts for most of the actual trotterization error.
This is illustrated in Figure~\ref{fig: fixed weights}.
In the left panel, we compare the different estimates of the trotterization error with the actual implementation error for a randomly generated $\SU(2)$ target, with $\calH$ given by the three Pauli matrices, $\sigma_X,\sigma_Y$, and $\sigma_Z$, and $\calC=\R^3$ (recall that $\calH$ forms a basis for $\su(2)$).
While the usual bound overestimates the observed approximation error by a significant margin, its transport-identity version in equation~\eqref{eq: fourier} remains remarkably close to it for all values of $N$ tested.
This is further confirmed by the right panel, where the bound in equation~\eqref{eq: fourier} exhibits an approximately linear relation with the observed $\varepsilon$, with slope close to $1$, for 100 randomly generated $\SU(2)$ targets.
We emphasize, however, that these experiments are intended to illustrate the error-interference mechanism rather than to compare constants with many-body bounds, where the relevant commutator norms and operator-norm prefactors may scale extensively with system size.

Similar cancellations occur in the nonzero-drift case, that is, whenever $\alpha\notin\calC$ and the minimum in equation~\eqref{eq: c as min} is strictly positive---for example, when the sets $\calC_j$ are restricted, as in the gate-synthesis case.
In particular, the leading linear drift contribution is
\begin{equation*}
    R_{\text{drift}} := \sum_{n=1}^N\Ad_{S^{1-n}}\Delta,
\end{equation*}
whose entries in the $H$-eigenbasis are
\begin{equation*}
    (R_{\text{drift}})_{ab} = e^{-i\frac{N-1}{2N}\omega_{ab}} \frac{\sin(\omega_{ab}/2)}{\sin(\omega_{ab}/2N)}
    \Delta_{ab}.
\end{equation*}
Thus, while the crude triangle inequality gives $\|R_{\text{drift}}\|\leq N\|\Delta\|$, geometric transport can make this contribution substantially smaller.
In the highly degenerate case, the geometric factor approaches $N$ and the expression agrees with the crude bound, whereas, in general, coherent cancellation may reduce it.

\section{Scheduled trotterization in commutator-closed control spaces}\label{sec: algorithm}

The fixed-weight solution described in the last section gives the scaling of the total trotterization error as $O(1/N)$.
Looking back at equation~\eqref{eq: full global error}, especially at the $R_{\text{int}}$ term, we see that, by appropriately adapting the schedule, we can make the accumulated error after each step bounded.
In particular, we propose \emph{artificially introducing drift terms} to cancel the error accumulated so far.

In this section, we transform this intuition into an algorithm.
For this purpose, define the \emph{accumulated internal error} at step $n$ as
\begin{equation}\label{eq: r int}
\begin{split}
R_{\text{int}}^{(n)} &:= \sum_{t=1}^n \Ad_{S^{1-t}}
\Bigg(
\Delta^{(t)}
+
\frac{1}{2}\sum_{i<j}
[c_i^{(t)}H_i,c_j^{(t)}H_j]
\\
&\qquad\qquad
-\frac{1}{2N}[H,\Delta^{(t)}]
\Bigg),
\end{split}
\end{equation}
where we let $R_{\text{int}}^{(0)}=0$.
Then, at each step $n$, we choose
\begin{equation}\label{eq: optimization c}
c^{(n+1)} \in \arg\min_{c\in\calC}
\left\| \sum_{j=1}^M(c_j-\alpha_j)H_j + \Ad_{S^n}R_{\text{int}}^{(n)} \right\|.
\end{equation}
Since
\begin{equation*}
\begin{aligned}
R_{\text{int}}^{(n+1)}
&=
R_{\text{int}}^{(n)}
+
\Ad_{S^{-n}}
\Bigg(
\Delta^{(n+1)}
\\
&\qquad
+
\frac{1}{2}\sum_{i<j}
\left[
c_i^{(n+1)}H_i,
c_j^{(n+1)}H_j
\right]
\\
&\qquad
-
\frac{1}{2N}
\left[
H,\Delta^{(n+1)}
\right]
\Bigg).
\end{aligned}
\end{equation*}
if the minimization in equation~\eqref{eq: optimization c} achieves exact cancellation,
\begin{equation*}
\Delta^{(n+1)} = -\Ad_{S^n}R_{\text{int}}^{(n)},
\end{equation*}
then
\begin{equation*}
\begin{aligned}
R_{\text{int}}^{(n+1)}
&=
\Ad_{S^{-n}}
\Bigg(
\frac{1}{2}\sum_{i<j}
\left[
c_i^{(n+1)}H_i,
c_j^{(n+1)}H_j
\right]
\\
&\qquad
-
\frac{1}{2N}
\left[
H,\Delta^{(n+1)}
\right]
\Bigg).
\end{aligned}
\end{equation*}
That is, the previously accumulated internal error is canceled, leaving only the new local contributions generated at the current step.
For later reference, this approach is formalized in Algorithm~\ref{alg: schedule}.

\begin{algorithm}[H]
\setcounter{AlgoLine}{0}
\caption{Schedule of trotterization with varying weights}
\label{alg: schedule}
\KwIn{A target unitary $U=\exp(iH)$.}
\KwRequire{A set of pulses $\calP$ and a value of $N\in\Z^+$.}

\KwOut{A schedule $(c^{(1)},\dots,c^{(N)})\in\calC^N$.}

$R_{\text{int}}^{(0)} \gets 0$\;

\For{$n \gets 1$ \KwTo $N$}{
$c^{(n)} \gets \arg\min_{c\in\calC} \left\| \sum_{j=1}^M(c_j-\alpha_j)H_j + \Ad_{S^{n-1}}R_{\text{int}}^{(n-1)} \right\|$

$\begin{aligned}
R_{\text{int}}^{(n)}
&\gets
R_{\text{int}}^{(n-1)}
+
\Ad_{S^{1-n}}
\Bigg(
\Delta^{(n)}
\\
&\qquad
+
\frac{1}{2}\sum_{i<j}
\left[
c_i^{(n)}H_i,
c_j^{(n)}H_j
\right]
\\
&\qquad
-
\frac{1}{2N}
\left[
H,\Delta^{(n)}
\right]
\Bigg)
\end{aligned}$\;
}
\end{algorithm}

\begin{remark}
The algorithm may be naturally extended to suppress the external error, but we do not attempt to do so here.
Instead, we guarantee that the drifts remain small at all steps, so their contribution is bounded.
\end{remark}

The algorithm's running time is governed by the number of steps, $N$, times the cost of the optimization step, which is usually the most expensive subroutine.
In practice, however, this is not as costly as it might initially appear: provided that $R_{\text{int}}^{(n-1)}$ is small, the desired correction to the usual trotterization weights is also small.
In other words, the solution is only \emph{slightly perturbed} from the usual choice of trotterization weights, while retaining the same pulse skeleton.

We note, however, that this cancellation strategy requires the commutators of elements of $\Span_{\R}\calH$ to be expressible as linear combinations of Hamiltonians in $\calH$; that is,
\begin{equation*}
    [\Span_{\R}\calH,\Span_{\R}\calH] \subseteq \Span_{\R}\calH.
\end{equation*}
Equivalently, $\Lie(\calH)=\Span_{\R}\calH$, where $\Lie(\calH)$ denotes the \emph{Lie closure} of $\calH$; see Appendix~\ref{sec: lie theory} for the definition.
Whenever this is the case, we say that the system has a \emph{commutator-closed control space}.
Throughout this section, we assume that $\Span_{\R}\calH$ is commutator-closed; an extension to the more general case is given in Section~\ref{sec: non lie closure}.

Besides the Lie-closure assumption, the global trotterization error of Algorithm~\ref{alg: schedule} also depends on the precise structure of $\calC$.
An ideal solution corresponds to setting, for all $n$,
\begin{equation*}
    \Delta^{(n)} = -\Ad_{S^{n-1}}R_{\text{int}}^{(n-1)}.
\end{equation*}
That is, at each step, we adjust the weights to cancel all previous internal error, leaving only the new local error introduced at the current step.
We call scenarios in which $\calC$ allows us to set
\begin{equation*}
    \left\| \Delta^{(n)} + \Ad_{S^{n-1}}R_{\text{int}}^{(n-1)} \right\| = 0
\end{equation*}
\emph{exact infinitesimal synthesis problems}.

On the other hand, when $\calC\neq\R^M$, it may be impossible to make the optimization objective in Algorithm~\ref{alg: schedule} exactly zero at every step; we call these \emph{imperfect infinitesimal synthesis problems}.
Still, even in the imperfect-synthesis case, the desired drift correction remains local, and we shall exploit this fact to estimate the global error.
We therefore divide the remainder of this section into these two infinitesimal-synthesis scenarios, establishing error bounds for each.

\subsection{Exact infinitesimal synthesis}

Given a set of available Hamiltonians, $\calH$, we define the \emph{synthesis map} as the linear operator
\begin{equation}\label{eq: synthesis map}
    \calB:\R^M\to\Span_{\R}\calH, \qquad \calB(c)=\sum_{j=1}^M c_jH_j.
\end{equation}
To quantify how changes in the synthesized Hamiltonian translate into changes in the pulse weights, we consider the \emph{control gain}
\begin{equation*}
    \kappa_{\calH} := \sup_{Y\in\Span_{\R}\calH,\;Y\neq 0} \frac{\min\{\|c\|:\calB(c)=Y\}}{\|Y\|},
\end{equation*}
where the norm in the numerator is the Euclidean norm in $\R^M$.
Thus, $\kappa_{\calH}$ bounds the size of the smallest change in pulse weights required to synthesize a given perturbation of the Hamiltonian.

\begin{theorem}[Exact infinitesimal synthesis]\label{th: exact synthesis}
    Suppose that $\Span_{\R}\calH$ is closed under commutators and that the fixed-weight pulse amplitudes satisfy
    \begin{equation*}
        \sum_{j=1}^M\|\alpha_jH_j\|=O(N^{-1}).
    \end{equation*}
    Assume that $\kappa_{\calH}$ is finite and independent of $N$, and that the admissible control set contains a neighborhood
    \begin{equation*}
        B\left(\alpha,\frac{K}{N^2}\right)\subseteq\calC
    \end{equation*}
    for some sufficiently large constant $K>0$ independent of $N$, where $B$ denotes the Euclidean ball in $\R^M$.
    Then, for sufficiently large $N$, Algorithm~\ref{alg: schedule} produces a trotterization schedule satisfying
    \begin{equation}\label{eq: exact synthesis error}
        \varepsilon=O(N^{-2}).
    \end{equation}
\end{theorem}

The origin of the quadratic improvement is particularly simple.
For ordinary first-order trotterization, each step generates an internal error of order $O(N^{-2})$, which may accumulate over $N$ steps to give the usual $O(N^{-1})$ global error.
Algorithm~\ref{alg: schedule} instead introduces an $O(N^{-2})$ perturbation of the next pulse weights, chosen to cancel the error accumulated so far.
Consequently, the internal error remains of order $O(N^{-2})$ throughout the evolution rather than accumulating linearly with $N$.
The remaining inter-step and higher-order contributions are also $O(N^{-2})$.
Complete finite-$N$ statement and proof are given in Appendix~\ref{sec: formalization}.

\begin{figure}
\centering
\includegraphics[width=0.5\textwidth]{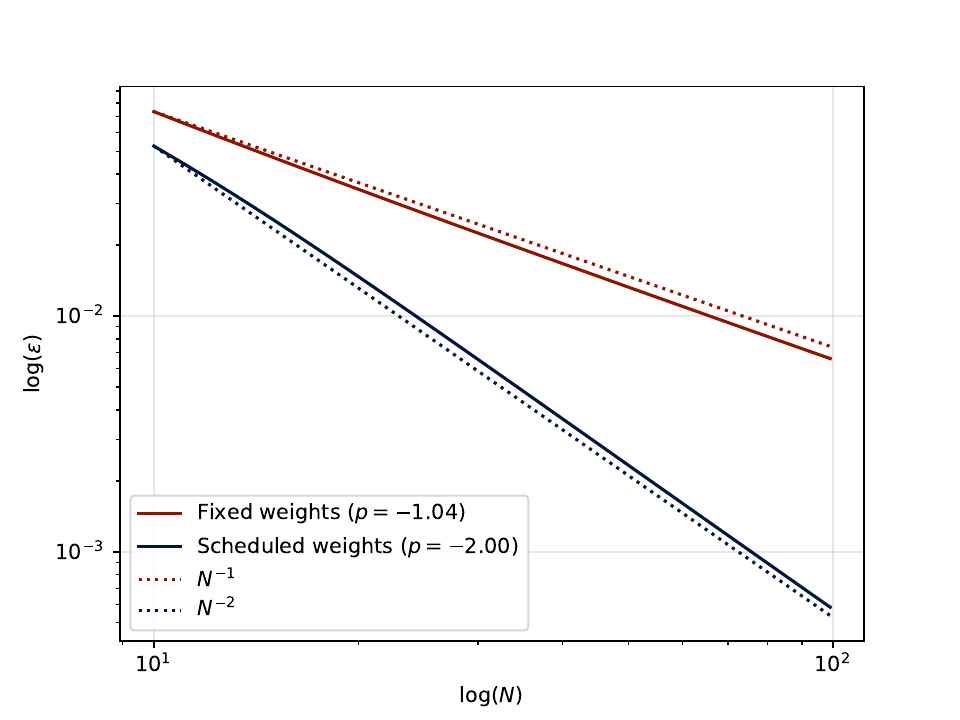}
\caption{Log-log plot of the observed trotterization error using the fixed- and scheduled-weight solutions. The fitted exponents ($p$), together with reference curves proportional to $N^{-1}$ and $N^{-2}$, are also shown. While the fixed-weight solution is consistent with an $O(N^{-1})$ decay, the scheduled-weight solution computed by Algorithm~\ref{alg: schedule} follows the $O(N^{-2})$ decay implied by Theorem~\ref{th: exact synthesis}.}
\label{fig: var vs fixed}
\end{figure}

We can empirically verify this improvement in the asymptotic dependence of the error on $N$.
Figure~\ref{fig: var vs fixed} shows the trotterization approximation error,
\begin{equation*}
    \varepsilon(N)=\|W(N)-U\|,
\end{equation*}
for a random $\SU(2)$ target as a function of $N$, on a log-log scale, where $W(N)$ denotes the $N$-step approximation.
Here, again, $\calH$ is the set of Pauli matrices and $\calC=\R^3$.
While the fitted line for the fixed-weight solution is consistent with a $1/N$ decay, variable-weight trotterization with the weights computed by Algorithm~\ref{alg: schedule} exhibits a behavior closer to $1/N^2$, in agreement with Theorem~\ref{th: exact synthesis}.
We observed similar fitted exponents across all tested random instances, although the initial transient depends on the target and therefore direct averaging of the curves is not very informative.

\subsection{Imperfect infinitesimal synthesis}\label{sec: imperfect}

When the admissible coefficients are restricted, the correction required by Algorithm~\ref{alg: schedule} may not be exactly implementable.
To quantify the corresponding loss of control resolution, for $r>0$ we define
\begin{equation}\label{eq: eta}
    \eta(X,r):= \sup_{\substack{Y\in\Span_{\R}\calH\\ \|Y-X\|\leq r}}
    \inf_{c\in\calC}\|\calB(c)-Y\|,
\end{equation}
where $\calB$ is the synthesis map of equation~\eqref{eq: synthesis map}.
Thus, $\eta(X,r)$ measures the largest synthesis error incurred when attempting to implement a Hamiltonian lying within a radius $r$ of $X$ using the admissible controls.
For instance, if $\calC=\Z^M$ and $\calB(\Z^M)$ forms a lattice in $\Span_{\R}\calH$, then $\eta(X,r)$ can be interpreted as a local covering radius of this lattice inside the ball $B(X,r)$~\cite{gruber1987geometry}.

The relevant regime for scheduled trotterization is when the available controls can approximate the $O(N^{-2})$ corrections generated by the algorithm with an error of the same order.
In this case, exact cancellation is no longer possible, but the accumulated internal error can still be confined to an $O(N^{-2})$ neighborhood.

\begin{theorem}[Imperfect infinitesimal synthesis]\label{th: imperfect synthesis}
    Suppose that $\Span_{\R}\calH$ is closed under commutators and that $\calH$ is linearly independent.
    Assume that, for some $\delta=O(N^{-1})$,
    \begin{equation*}
        \sum_{j=1}^M\|\alpha_jH_j\|\leq \frac{\delta}{2},
    \end{equation*}
    and that $\kappa_{\calH}$ is finite and independent of $N$.

    If there exists a radius $r=O(N^{-2})$ such that
    \begin{equation}\label{eq: imperfect invariant radius}
        \eta(\calB(\alpha),r)
        +
        \frac{\delta^2}{2}
        +
        \delta\left(r+\eta(\calB(\alpha),r)\right)
        \leq r,
    \end{equation}
    then, for sufficiently large $N$,
    Algorithm~\ref{alg: schedule} produces a trotterization schedule satisfying
    \begin{equation}\label{eq: imperfect synthesis error general}
        \varepsilon=O(N^{-2}).
    \end{equation}
\end{theorem}

Condition~\eqref{eq: imperfect invariant radius} has a simple interpretation.
The synthesis error introduced by the restricted controls, together with the new local Trotter error and their interaction, must fit inside the same radius $r$ at every step.
When $r=O(N^{-2})$, this prevents the internal error from accumulating beyond the scale already generated by a single Trotter step.

Note that the exact-synthesis case is recovered when
\begin{equation*}
    \eta(\calB(\alpha),r)=0.
\end{equation*}
Again, a complete finite-$N$ statement and its proof are given in Appendix~\ref{sec: formalization}.

We illustrate the imperfect-synthesis case by applying it to trotterization-based methods for gate synthesis.
Recall that a gate $G_j$ can be modeled as the pulse $G_j=\exp(iH_j)$, whereas repeated applications of the gate or its inverse are described by letting $\calC_j=\Z$; as usual, we assume that the inverse gates can also be implemented~\cite{dawson2005solovay}.
Following~\cite{ennes2026removingonlineexponentialnet}, we call an implementable set of gates $G_1,\dots,G_M$ (including their inverses) an \emph{exponential basis} if
\begin{equation*}
    \calH=\{H_j=\log G_j\mid 1\leq j\leq M\}
\end{equation*}
forms a basis for $\su(d)$---in particular, $M=d^2-1$.
Provided that the maximum norm of the available Hamiltonians,
\begin{equation*}
    h_{\max}:=\max\{\|H_j\|\mid H_j\in\calH\},
\end{equation*}
is small, the exponential basis may be used as the set of pulses for trotterization-based unitary synthesis.
In particular, any $Y\in\su(d)$ may be approximated by $\calB(c)$, with $c\in\Z^M$, within error at most $Mh_{\max}/2$, implying that
\begin{equation*}
    \eta(\calB(\alpha),r)\leq \frac{Mh_{\max}}{2}
\end{equation*}
for all $r>0$.
Therefore, referring to equation~\eqref{eq: imperfect synthesis error general}, a sufficient scaling condition for remaining in the expected second-order regime in $\delta$ is
\begin{equation*}
    Mh_{\max}=O(\delta^2),
\end{equation*}
where $\delta$ is constrained by the assumptions in equations~\eqref{eq: hypothesis alpha imperfect 1} and~\eqref{eq: hypothesis alpha imperfect 2}.
Thus, for a family of exponential bases satisfying this condition, if the remaining hypotheses of Theorem~\ref{th: imperfect synthesis} hold and $\delta=\Theta(1/N)$, the corresponding error scales as $O(1/N^2)$.

\begin{figure}[t]
    \centering
    \includegraphics[width=0.48\textwidth]{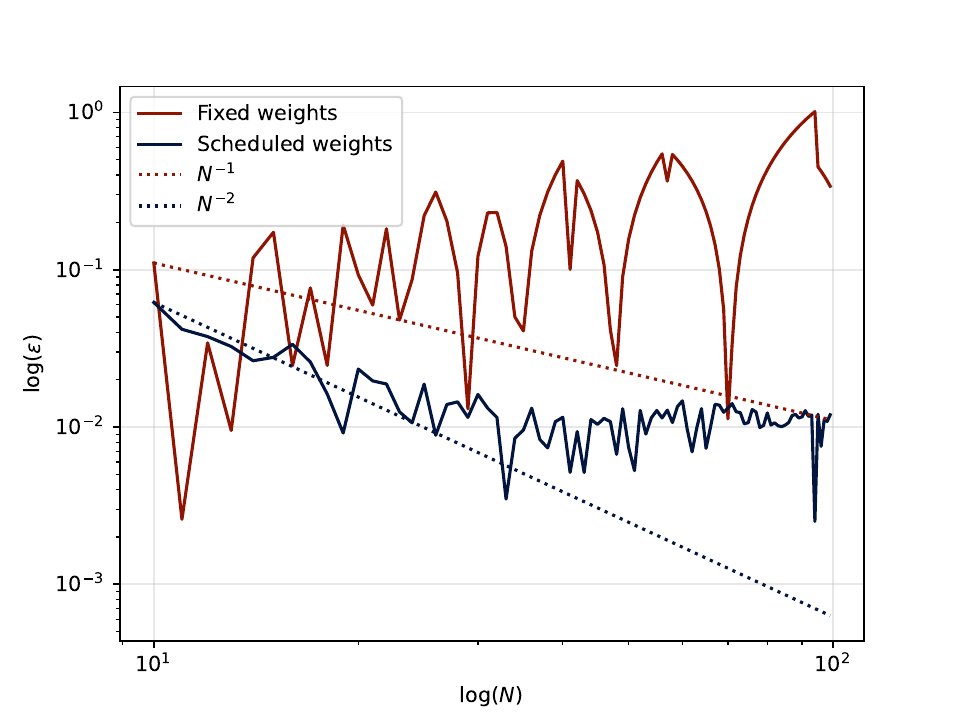}
    \hfill
    \includegraphics[width=0.48\textwidth]{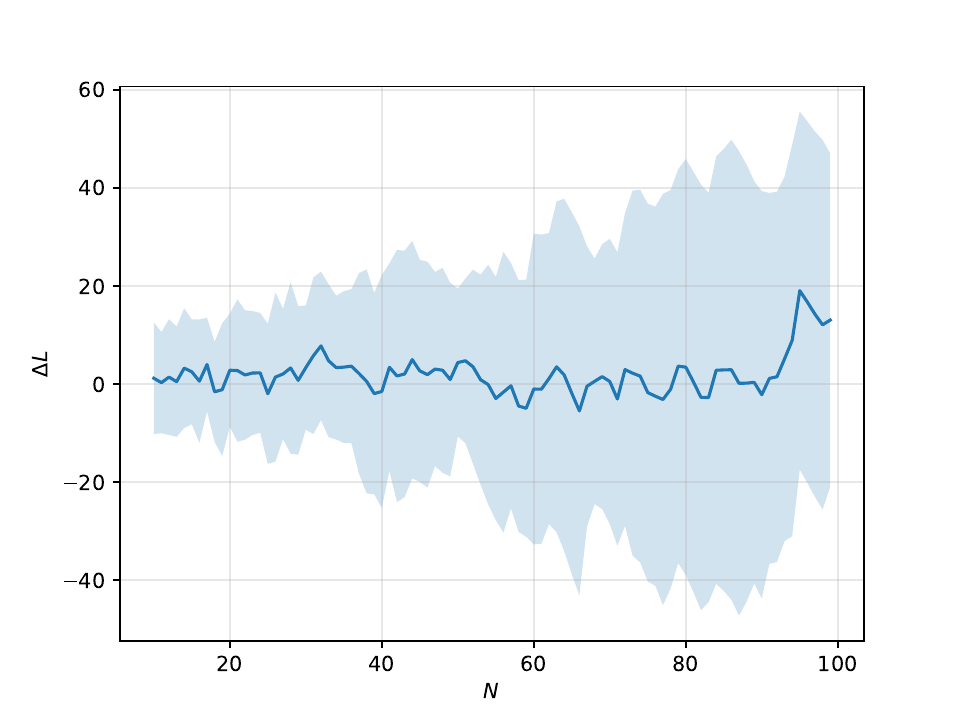}
    \caption{\textbf{Left:} Log-log plot of the observed trotterization error using the fixed- and scheduled-weight solutions for $\calC=\Z^3$, where $\calH$ was taken to be small multiples of the three Pauli matrices and the target is a random $\SU(2)$ unitary. Reference curves proportional to $N^{-1}$ and $N^{-2}$ are also shown. \textbf{Right:} The difference $\Delta L$ between the number of gates in the scheduled- and fixed-weight solutions for trotterization-based gate synthesis. Statistics were taken over 30 randomly generated $\SU(2)$ targets, and the filled region corresponds to one standard deviation around the average.}
    \label{fig: gates}
\end{figure}

We numerically illustrate this gate-synthesis scenario in Figure~\ref{fig: gates}.
Here, we take
\begin{equation*}
    \calH=\{k\cdot\sigma_x,k\cdot\sigma_y,k\cdot\sigma_z\},
\end{equation*}
where $\sigma_i$ are the Pauli matrices and $k$ is a small normalization constant, which we set to $0.01$.
We use Algorithm~\ref{alg: schedule} to find an imperfect-synthesis trotterization scheme using these pulses for a randomly generated $\SU(2)$ target.
Instead of exactly solving the optimization in equation~\eqref{eq: optimization c}, we simply round the real-valued solution; that is, we let
\begin{equation*}
\begin{aligned}
c^{(n+1)}
&\gets
\operatorname{round}\Bigg(
\arg\min_{c\in\R^3}
\Bigg\|
\sum_{j=1}^3
(c_j-\alpha_j)H_j
\\
&\qquad\qquad
+
\Ad_{S^n}R_{\text{int}}^{(n)}
\Bigg\|
\Bigg).
\end{aligned}
\end{equation*}
Because $k$ is small, the discretization error introduced by this rounding remains small over the finite regime considered.
As the left panel of the figure shows, the variable-weight solution gives smaller errors than the fixed-weight solution for most values of $N$.
Because of discretization, however, the relation between $N$ and $\varepsilon$ is too noisy to exhibit a clear power law in $1/N$, especially when $N\sim1/k$.

An interesting follow-up question is whether scheduled trotterization requires more gates for its implementation than the fixed-weight solution.
For integer coefficients, we define the gate counts as
\begin{equation*}
    L_{\text{schedule}} := \sum_{n=1}^N\sum_{j=1}^M|c_j^{(n)}|, \quad L_{\text{fixed}} := N\sum_{j=1}^M|\operatorname{round}(\alpha_j)|,
\end{equation*}
and let
\begin{equation*}
    \Delta L:=L_{\text{schedule}}-L_{\text{fixed}}.
\end{equation*}
Since $c_j^{(n)}$ is adjusted at each step around the usual fixed-weight choice, some scheduled coefficients may be larger in absolute value than the corresponding fixed weights, while others may be smaller.
It is therefore plausible that these changes partially balance when accumulated over the whole schedule, although there is no \emph{a priori} reason for exact cancellation.
The right panel of Figure~\ref{fig: gates} indicates that this is indeed the case in our numerical experiments.
Here, we show the value of $\Delta L$ estimated over 30 random targets in $\SU(2)$ for varying choices of $N$.
While the standard deviation appears to grow with $N$, the average remains close to $0$.
We do not attempt to characterize the precise stochastic process here, nor do we investigate how it may depend on the dimension $d$ and the scale $k$, but these experiments suggest that scheduled trotterization for gate synthesis may achieve improved error scaling without substantially increasing the average number of gates.

\begin{remark}
In~\cite{ennes2026removingonlineexponentialnet}, it is argued that if an exponential basis can be used to approximate any unitary $U\in\SU(d)$ within some universal constant error $\varepsilon_0$, then the corresponding gate set is universal for quantum computing.
In particular, provided that $h_{\max}$ is sufficiently small, $G_1,\dots,G_M$ will be universal.
Using fixed-weight trotterization schemes, the required scale is estimated as $h_{\max}=O(\varepsilon_0^2)$.
The discussion of this section suggests that, if scheduled trotterization is allowed, this requirement might be weakened to $h_{\max}=O(\varepsilon_0)$, while the asymptotic dependence on $M$ appears to remain similar (\ie, of order $O(M^{-3})\sim d^{-6}$).
\end{remark}

\section{Beyond commutator-closed control spaces}\label{sec: non lie closure}

A key requirement for the cancellation mechanism of Algorithm~\ref{alg: schedule} is that
\begin{equation*}
    \Ad_{S^{n-1}}R_{\text{int}}^{(n-1)}
    \in\Span_{\R}\calH.
\end{equation*}
When this fails, as may occur when $\Lie(\Span_{\R}\calH)\neq\Span_{\R}\calH$, the transported residual can contain a component outside $\Span_{\R}\calH$ that cannot be canceled by any choice of drift.
Although the algorithm may still be applied in this case, its error is no longer guaranteed to remain small enough to yield an improvement.

More generally, within the present framework, it does not seem reasonable to expect the fixed-skeleton cancellation strategy to directly address this lack of closure.
At each step, the error contains commutators of $\calH$-terms, modulo adjoint actions, which may generate directions outside $\Span_{\R}\calH$.
One could in principle allow such components to accumulate in the hope that they later return close to $\Span_{\R}\calH$, but this would provide no stable cancellation guarantee.
We therefore do not pursue such fixed-skeleton heuristics here.

If, however, we are willing to modify the pulse skeleton, the ideas of the previous section can still be applied after enlarging the available control space.
The construction follows the proof of the Lie-algebra rank condition theorem~\cite{d2021introduction, jurdjevic1972control}: we enlarge $\calH$ until its real span equals its Lie closure.
Importantly, the additional Hamiltonians introduced below remain implementable using products of pulses generated by the original Hamiltonians.
Indeed, if $H_{\texttt{new}} =  \Ad_{\exp(i\tau H_i)}H_j$, then
\begin{equation*}
    \exp(icH_{\texttt{new}})
    =
    \exp(i\tau H_i)
    \exp(icH_j)
    \exp(-i\tau H_i).
\end{equation*}

The procedure is described in Algorithm~\ref{alg: expand}.
Here, we assume continuously tunable controls, so that the conjugation times $\tau$ appearing in the construction are implementable.
Its correctness is established in Proposition~\ref{prop: expansion}.
Since the proof is very similar to that of Lemma~1 in Appendix~D of~\cite{d2021introduction}, we defer it to Appendix~\ref{app: proof expansion}.

\begin{proposition}\label{prop: expansion}
Suppose that Algorithm~\ref{alg: expand} is performed in exact arithmetic (\eg, with $\texttt{tol}=0$).
Then Algorithm~\ref{alg: expand} terminates after
\begin{equation*}
    \calM
    =
    \dim\Lie(\Span_{\R}\calH)
    -
    \dim\Span_{\R}\calH
\end{equation*}
successful augmentations, where $\calM$ is the synthesis defect of the initial set $\calH$.
\end{proposition}

\begin{algorithm}[H]
\setcounter{AlgoLine}{0}
\caption{Extend $\calH$ to its Lie closure}
\label{alg: expand}
\KwIn{A list of available Hamiltonians $\calH=\{H_1,\dots,H_M\}$.}
\KwRequire{A tolerance threshold $\texttt{tol}$.}
\KwOut{An updated list of Hamiltonians
$\calH=\{H_1,\dots,H_M,H_{M+1},\dots,H_{\widetilde M}\}$
satisfying $\Lie(\Span_{\R}\calH)=\Span_{\R}\calH$.}

\While{$\Lie(\Span_{\R}\calH)\neq\Span_{\R}\calH$}{
choose $H_i,H_j\in\calH$ such that
$[H_i,H_j]\notin\Span_{\R}\calH$
\tcp{find, e.g., by row reduction}

$\Pi\gets$ Hilbert-Schmidt orthogonal projection onto $\Span_{\R}\calH$\;

\For{$k\gets0,1,\dots$}{
$\tau\gets 2^{-k}$\;

$H_{\texttt{new}}\gets\Ad_{\exp(i\tau H_i)}H_j$\;

\If{$\|(I-\Pi)H_{\texttt{new}}\|>\texttt{tol}$}{
$\calH.\texttt{append}(H_{\texttt{new}})$\;

\texttt{break}\;
}
}
}
\end{algorithm}

Algorithm~\ref{alg: expand} can therefore be viewed as a preprocessing step whenever $\Span_{\R}\calH$ is not closed under commutators.
We first enlarge $\calH$ to a new set $\calH'$ and then use $\calH'$ as the pulse Hamiltonians for Algorithm~\ref{alg: schedule}.
The results of the previous section can then be applied, subject to the same assumptions on the admissible controls and infinitesimal synthesis.

We once more stress, however, that the enlarged pulse set necessarily changes the trotterization skeleton produced by this construction relative to the original fixed-weight solution.
Moreover, implementing the additional Hamiltonians requires conjugation by other available pulses, so the total number of applied pulses may exceed the usual $NM$.

\section{Discussion}
We introduced scheduled trotterization as a variable-weight alternative to fixed-weight product formulas.
Instead of repeating the same local approximation at every step, the method described by Algorithm~\ref{alg: schedule} chooses a schedule of admissible weights so that the local synthesis errors cancel after being transported along the ideal evolution.
The basic mechanism is captured by the transport identity which rewrites the global endpoint error as a product of Lie-algebraic defects expressed in a common moving frame.

This viewpoint separates two sources of error.
The first is the internal error, coming from the local BCH expansion of each constrained step.
The second is the external error, coming from the BCH expansion of the transported defects across different steps. 
This natural split gives a useful way to design schedules: rather than minimizing each local error independently, one may choose local errors whose transported sum remains small.
In particular, our scheduling rule keeps the leading internal residual bounded throughout the evolution.

For exact infinitesimal synthesis, we proved that this mechanism improves the global endpoint scaling from the usual fixed-weight $O(N^{-1})$ behavior to an $O(N^{-2})$ error bound, under the same local BCH assumptions.
Similar framework also applies to imperfect infinitesimal synthesis, where the local covering parameter $\eta$ measures how well the admissible coefficient set approximates the desired infinitesimal corrections.
In this case, the error bound depends on the radius of an invariant residual ball, giving a perturbative criterion under which the scheduled construction still improves over naive fixed-weight repetition.

The numerical experiments support this theoretical picture. 
For random $\SU(2)$ targets with Pauli generators and continuous admissible weights, the fixed-weight product formula exhibits a decay consistent with $1/N$, while the scheduled weights constructed by Algorithm~\ref{alg: schedule} exhibit a decay closer to $1/N^2$.

More broadly, the results suggest that Trotter errors should not only be viewed as quantities to be bounded locally.
In coherent product-formula evolutions, they can also be treated as schedulable objects whose transported accumulation may be cancelled at the endpoint.
This gives a bridge between product-formula analysis, Lie-algebraic control, and constrained gate synthesis, and suggests several directions for future work.

Of these, an important extension of Algorithm~\ref{alg: schedule} is its specialization to linear combinations of sparse Pauli hermitians.
This setting is particularly relevant for many-body simulation and quantum chemistry, where the target Hamiltonians and/or the available Hamiltonians are most often assumed assumed to be of this form.
While in a Pauli basis, coefficient recovery is simple because Pauli strings are orthogonal, computing the transported residuals $\Ad_{S^{1-n}}R$ without an uncontrolled growth of the active Pauli support does not seem trivial.

\begin{acknowledgments}
This work has been partially supported by the ANR project ANR-20-CE48-0007 (AlgoKnot) and the project ANR-15-IDEX-0001 (UCA JEDI). It has also been supported by the French government, through the France 2030 investment plan managed by the Agence Nationale de la Recherche, as part of the ``UCA DS4H'' project, reference ANR-17-EURE-0004.

The author would like to thank Chih-Kang Huang for his inputs on an earlier version of this work.
\end{acknowledgments}
\bibliography{apssamp}

\clearpage
\onecolumngrid
\appendix

\section{Review of Lie theory}\label{sec: lie theory}

In the usual model of finite-dimensional quantum mechanics, pure physical states are represented by normalized vectors in a finite-dimensional complex Hilbert space, $\C^d$ (up to a global phase), while their evolution is given by a \emph{special unitary operator} $U$.
That is, $U$ is a $d\times d$ complex matrix with $U^\dagger=U^{-1}$ and $\det U=+1$.
The set of $d\times d$ special unitary operators, $\SU(d)$, naturally forms a group under ordinary matrix multiplication.
In fact, since the product and inversion maps of $\SU(d)$ are smooth, it has the structure of a (real) \emph{Lie group} of dimension $d^2-1$.
In particular, $\SU(d)$ is a \emph{compact Lie group}.

Much of the local structure of $\SU(d)$ is encoded in its tangent space at the identity, $T_I\SU(d)$.
In the standard convention, this tangent space consists of skew-Hermitian traceless matrices.
However, throughout this paper, we identify it (through multiplication by $i$) with the vector space of Hermitian ($X^\dagger=X$) and traceless ($\tr X=0$) $d\times d$ complex matrices, which we denote by $\su(d)$.
Besides naturally forming a real vector space of dimension $d^2-1$, $\su(d)$, when endowed with the real-bilinear and antisymmetric Lie bracket
\begin{equation*}
    [X,Y]:=i(XY-YX),
\end{equation*}
is a \emph{Lie algebra}~\cite{erdmann2006introduction, hall2013lie}, which we shall refer to as the Lie algebra of $\SU(d)$.

Given a subset $\calH$ of $\su(d)$, we define its \emph{Lie closure}, $\Lie(\calH)$, as the smallest Lie subalgebra of $\su(d)$ that contains $\Span_{\R}\calH$.
For example, if $\calH=\{\sigma_X,\sigma_Z\}\subseteq\su(2)$ consists of the $X$ and $Z$ Pauli matrices, then
\begin{equation*}
    \Lie(\calH)=\Span_{\R}\{\sigma_X,\sigma_Z,\sigma_Y\},
\end{equation*}
where $\sigma_Y$ is the $Y$ Pauli matrix.
The \emph{synthesis defect} of $\calH\subseteq\su(d)$ is the nonnegative integer
\begin{equation*}
    \calM := \dim\Lie(\calH) - \dim\Span_{\R}\calH.
\end{equation*}
In the case $\calH=\{\sigma_X,\sigma_Z\}$, for example, it equals $1$.

For any $X\in\su(d)$, its complex matrix exponential
\begin{equation}\label{eq: matrix exponential}
    \exp(iX):=\sum_{n=0}^\infty\frac{(iX)^n}{n!}
\end{equation}
is an element of $\SU(d)$.
Thus, under our Hermitian convention, the exponential map
\begin{equation*}
    X\mapsto\exp(iX)
\end{equation*}
is a smooth map from $\su(d)$ to $\SU(d)$, although it is not a diffeomorphism.
Still, we can always choose a (non-smooth) right inverse of the exponential map,
$\log:\SU(d)\to\su(d)$, such that
\begin{equation*}
    \exp(i\log U)=U
    \qquad\text{and}\qquad
    \|\log U\|\leq 2\pi
\end{equation*}
for all $U\in\SU(d)$~\cite{ennes2026removingonlineexponentialnet}.
Here and throughout, $\|\cdot\|$ denotes the \emph{operator norm},
\begin{equation}\label{eq: operator norm}
    \|A\| := \sup_{v\neq0}\frac{\|Av\|}{\|v\|},
\end{equation}
where $v\in\C^d$ and the norm on the right-hand side of equation~\eqref{eq: operator norm} is the usual Euclidean norm on $\C^d$.

\begin{remark}
For normal matrices, which include unitary and Hermitian matrices, the operator norm equals the largest absolute value of their eigenvalues (which, for unitaries, is always $1$).
\end{remark}

In the main text, we use some important properties of matrix exponentials and logarithms.
First, we note that
\begin{equation}\label{eq: conjugation}
    U\exp(iX)U^{-1} = \exp(iUXU^{-1})
\end{equation}
for any $X\in\su(d)$ and $U\in\SU(d)$.
For this to be well-defined, one needs $UXU^{-1}\in\su(d)$ for all choices of $X$ and $U$.
This is indeed the case, since the map
\begin{equation*}
    \Ad_U(X):=UXU^{-1}
\end{equation*}
is the well-known \emph{adjoint representation} of the Lie group $\SU(d)$ on its Lie algebra $\su(d)$.
With this notation, equation~\eqref{eq: conjugation} becomes
\begin{equation*}
    U\exp(iX)U^{-1} = \exp(i\Ad_UX).
\end{equation*}

Second, we will need some bounds on the operator norms involved.
These are summarized in the lemma below, whose proof can be found in~\cite{ennes2026removingonlineexponentialnet}.
\begin{lemma}\label{lm: properties operator norm}
Suppose that $U,V\in\SU(d)$ are unitary, $X,Y\in\su(d)$ are Hermitian, and $A$ is a $d\times d$ complex matrix.
Then
\begin{enumerate}[label=(\alph*)]
    \item\label{it: unitary invariance}
    $\|UAV\|=\|A\|$;
    \item\label{it: exponent approximation}
    $\|\exp(iX)-\exp(iY)\|\leq\|X-Y\|$.
\end{enumerate}
\end{lemma}

\section{Proof of Theorem~\ref{th: variable trotter error}}\label{sec: proof error bound}

\begin{proof}[Proof of Theorem~\ref{th: variable trotter error}]
Before proceeding with the main proof, we first record the norms of the quantities of interest.
By assumption,
\begin{equation*}
\left\|\frac{H}{N}\right\|\leq\delta, \qquad \sum_{j=1}^M\|c_j^{(n)}H_j\|\leq\delta
\end{equation*}
for every $1\leq n\leq N$.
Hence
\begin{equation*}
\left\|\Delta^{(n)}\right\| = \left\| \sum_{j=1}^M c_j^{(n)}H_j-\frac{H}{N} \right\| \leq \sum_{j=1}^M\|c_j^{(n)}H_j\| + \left\|\frac{H}{N}\right\| \leq 2\delta.
\end{equation*}
Moreover, the total norm of the exponents entering each local BCH expansion is at most $2\delta\leq\delta_0$, so the expansion applies to every local defect.

We now proceed to the main argument.
As described in the main text, its core consists of two applications of BCH to the transport identity: one accounting for errors between different steps, corresponding to $R_{\text{ext}}$, and the other for errors within a single step, corresponding to $R_{\text{int}}$.
By the transport identity,
\begin{equation*}
    S^{-N}\overleftarrow{\prod_{t=1}^N}T^{(t)} = \overleftarrow{\prod_{t=1}^N} \exp\left(i\Ad_{S^{1-t}}X^{(t)}\right).
\end{equation*}
Therefore, using unitary invariance of the norm,
\begin{equation}\label{eq: interstep bch}
\begin{split}
    \varepsilon &=\|W-U\|\\
    &= \left\| \overleftarrow{\prod_{t=1}^N} \exp\left(i\Ad_{S^{1-t}}X^{(t)}\right) - I \right\|\\
    &\leq \left\| \log\left( \overleftarrow{\prod_{n=1}^N} \exp\left(i\Ad_{S^{1-n}}X^{(n)}\right) \right) \right\|\\
    &\leq \left\|\sum_{n=1}^N\Ad_{S^{1-n}}X^{(n)} + \frac{1}{2}\sum_{m<n} \left[\Ad_{S^{1-n}}X^{(n)},\Ad_{S^{1-m}}X^{(m)}\right]\right\| + \|R_3^1\|.
\end{split}
\end{equation}
Here BCH applies because
\begin{equation*}
    \sum_{n=1}^N \left\|\Ad_{S^{1-n}}X^{(n)}\right\| = \sum_{n=1}^N\|X^{(n)}\| = \sigma \leq\delta_0.
\end{equation*}
Moreover,
\begin{equation*}
    \|R_3^1\|\leq C\sigma^3.
\end{equation*}

Similarly, applying BCH to the local defect $S^{-1}T^{(n)}$, we get
\begin{equation}\label{eq: local defect expansion}
\begin{split}
    X^{(n)} &= \log(S^{-1}T^{(n)})\\
    &=\log\left( \exp\left(-i\frac{H}{N}\right) \overrightarrow{\prod_{j=1}^M} \exp\left(ic_j^{(n)}H_j\right) \right)\\
    &= \sum_{j=1}^M c_j^{(n)}H_j-\frac{H}{N} + \frac{1}{2}\sum_{i<j} \left[c_i^{(n)}H_i,c_j^{(n)}H_j\right] - \frac{1}{2}\sum_{j=1}^M \left[\frac{H}{N},c_j^{(n)}H_j\right] + R_3^{2,n}\\
    &= \Delta^{(n)} + \frac{1}{2}\sum_{i<j} \left[c_i^{(n)}H_i,c_j^{(n)}H_j\right] - \frac{1}{2N} \left[H,\Delta^{(n)}\right] + R_3^{2,n},
\end{split}
\end{equation}
where in the last line we used the bilinearity of the commutator and $[H,H/N]=0$.
The local BCH remainder satisfies
\begin{equation*}
    \|R_3^{2,n}\|\leq C\delta^3
\end{equation*}
uniformly in $n$, after absorbing universal constants into $C$.

Substituting equation~\eqref{eq: local defect expansion} into equation~\eqref{eq: interstep bch}, the linear contribution is
\begin{equation*}
    R_{\text{int}} = \sum_{n=1}^N \Ad_{S^{1-n}} \left(\Delta^{(n)} + \frac{1}{2}\sum_{i<j} \left[c_i^{(n)}H_i,c_j^{(n)}H_j\right] - \frac{1}{2N} \left[H,\Delta^{(n)}\right]\right).
\end{equation*}
The leading external quadratic contribution is
\begin{equation*}
R_{\text{ext}} = \frac{1}{2}\sum_{m<n}\left[\Ad_{S^{1-n}}\Delta^{(n)}, \Ad_{S^{1-m}}\Delta^{(m)}\right].
\end{equation*}
We collect all remaining terms into $R_3$.

We now estimate these remaining terms.
The accumulated local BCH remainders contribute
\begin{equation*}
    \left\| \sum_{n=1}^N\Ad_{S^{1-n}}R_3^{2,n} \right\| \leq CN\delta^3,
\end{equation*}
while the external BCH remainder contributes
\begin{equation*}
    \|R_3^1\|\leq C\sigma^3.
\end{equation*}

It remains to estimate the terms omitted when replacing the full external commutator
\begin{equation*}
    \frac{1}{2}\sum_{m<n} \left[\Ad_{S^{1-n}}X^{(n)},\Ad_{S^{1-m}}X^{(m)}\right]
\end{equation*}
by the $\Delta$--$\Delta$ contribution defining $R_{\text{ext}}$.
Write
\begin{equation*}
    X^{(n)} = \Delta^{(n)}+Q^{(n)}+R_3^{2,n},
\end{equation*}
where
\begin{equation*}
    Q^{(n)} := \frac{1}{2}\sum_{i<j} [c_i^{(n)}H_i,c_j^{(n)}H_j] - \frac{1}{2N}[H,\Delta^{(n)}].
\end{equation*}
Set
\begin{equation*}
    \beta:=\max_{1\leq n\leq N}\|\Delta^{(n)}\|,
\end{equation*}
and note that $\beta\leq2\delta$.
Then, by the local-size assumptions,
\begin{equation*}
    \left\| \frac{1}{2}\sum_{i<j} [c_i^{(n)}H_i,c_j^{(n)}H_j] \right\| \leq\frac{\delta^2}{2},
\end{equation*}
and
\begin{equation*}
    \frac{1}{2N}\|[H,\Delta^{(n)}]\| \leq \left\|\frac{H}{N}\right\| \|\Delta^{(n)}\| \leq \delta\beta.
\end{equation*}
Since $\beta\leq2\delta$, after increasing $C$ we have
\begin{equation*}
    \|Q^{(n)}\|\leq C\delta^2.
\end{equation*}

Therefore, every omitted external commutator contains at least one factor $Q^{(n)}$ or one local remainder $R_3^{2,n}$.
The terms containing one $\Delta$ factor and one $Q$ factor contribute at most
\begin{equation*}
    CN^2\beta\delta^2,
\end{equation*}
while the terms containing two $Q$ factors contribute at most
\begin{equation*}
    CN^2\delta^4.
\end{equation*}
The terms involving at least one local remainder $R_3^{2,n}$ are smaller and can be absorbed into the same bound, using
$\|R_3^{2,n}\|\leq C\delta^3$,
$\beta\leq2\delta$, and $\delta\leq1$.

Combining these estimates, after increasing $C$ if necessary, we obtain
\begin{equation*}
    \|R_3\|  \leq C\left(N\delta^3 + N^2\beta\delta^2 + N^2\delta^4 + \sigma^3 \right).
\end{equation*}
Therefore,
\begin{equation*}
    \varepsilon \leq \|R_{\text{int}}\| + \|R_{\text{ext}}\| + C\left( N\delta^3 + N^2\beta\delta^2 + N^2\delta^4 + \sigma^3\right).
\end{equation*}
\end{proof}

\section{Formalization of Theorems~\ref{th: exact synthesis} and~\ref{th: imperfect synthesis}}\label{sec: formalization}

We state here finite-$N$ versions of Theorems~\ref{th: exact synthesis} and~\ref{th: imperfect synthesis} and prove them.
We start with the exact-synthesis case.

\begin{theorem}[Exact infinitesimal synthesis (finite $N$)]
    Suppose that $\Lie(\calH)=\Span_{\R}\calH$ and
    \begin{equation}\label{eq: hypothesis alpha exact 1}
         CN\delta^2\leq\delta_0,
         \qquad
         2\delta\leq\delta_0,
         \qquad\text{and}\qquad
         \sum_{j=1}^M\|\alpha_jH_j\|\leq\frac{\delta}{2},
    \end{equation}
    where $C$ and $\delta_0$ are sufficiently large and sufficiently small universal constants, respectively.
    Suppose moreover that the set of available Hamiltonians $\calH$ satisfies
    \begin{equation}\label{eq: hypothesis alpha exact 2}
        \kappa_{\calH}
        \leq
        \left(
        2\delta\sqrt{\sum_{j=1}^M\|H_j\|^2}
        \right)^{-1}.
    \end{equation}
    Define
    \begin{equation*}
        \rho:=\kappa_{\calH}\delta^2
    \end{equation*}
    and let
    \begin{equation*}
        B(\alpha,\rho) := \{c\in\R^M:\|c-\alpha\|\leq\rho\}
    \end{equation*}
    be the closed Euclidean ball of radius $\rho$ centered at $\alpha$.
    If $B(\alpha,\rho)\subseteq\calC$, then Algorithm~\ref{alg: schedule} produces a trotterization schedule whose total global error satisfies
    \begin{equation*}
        \varepsilon \leq K_{\text{exact}} \left( \delta^2+N^2\delta^4+N\delta^3 \right),
    \end{equation*}
    where $K_{\text{exact}}$ is a universal constant.
\end{theorem}

\begin{proof}
    We start by showing that, for all $1\leq n\leq N$, $\|R_{\text{int}}^{(n)}\|\leq\delta^2$ and that the choice
    \begin{equation*}
        \Delta^{(n)} = -\Ad_{S^{n-1}}R_{\text{int}}^{(n-1)}
    \end{equation*}
    can be obtained using a coefficient vector in $B(\alpha,\rho)$.

    Assume inductively that $\|R_{\text{int}}^{(n-1)}\|\leq\delta^2$.
    Since $H\in\Span_{\R}\calH$ and $\Lie(\calH)=\Span_{\R}\calH$, the space $\Span_{\R}\calH$ is invariant under the linear operator $[H,\cdot]$ and therefore under
    \begin{equation*}
        \Ad_{\exp(itH)}=\exp(t[H,\cdot])
    \end{equation*}
    for every $t\in\R$.
    It follows inductively from equation~\eqref{eq: r int} that
    \begin{equation*}
        R_{\text{int}}^{(n-1)}
        \in\Span_{\R}\calH.
    \end{equation*}
    Therefore, there exists $u^{(n)}\in\R^M$ such that
    \begin{equation*}
        \calB(u^{(n)}) = -\Ad_{S^{n-1}}R_{\text{int}}^{(n-1)}.
    \end{equation*}
    Moreover, we may choose $u^{(n)}$ so that
    \begin{equation*}
        \|u^{(n)}\| \leq \kappa_{\calH}  \|\calB(u^{(n)})\| \leq \kappa_{\calH}\delta^2.
    \end{equation*}
    Therefore, letting
    \begin{equation*}
        c^{(n)}=\alpha+u^{(n)},
    \end{equation*}
    we have $c^{(n)}\in B(\alpha,\rho)\subseteq\calC$ and
    \begin{equation*}
        \Delta^{(n)} = \calB(u^{(n)}) = -\Ad_{S^{n-1}}R_{\text{int}}^{(n-1)}.
    \end{equation*}
    Furthermore,
    \begin{equation}\label{eq: bound r int exact}
    \begin{split}
        \sum_{j=1}^M\|c_j^{(n)}H_j\| &\leq \sum_{j=1}^M\|\alpha_jH_j\| + \sum_{j=1}^M\|u_j^{(n)}H_j\|\\
        &\leq \frac{\delta}{2} + \sqrt{\sum_{j=1}^M|u_j^{(n)}|^2} \sqrt{\sum_{j=1}^M\|H_j\|^2}\\
        &\leq\delta,
    \end{split}
    \end{equation}
    by equations~\eqref{eq: hypothesis alpha exact 1} and~\eqref{eq: hypothesis alpha exact 2}.

    By the choice of $\Delta^{(n)}$, its contribution cancels the previously accumulated internal error, and hence
    \begin{equation*}
    \begin{split}
       \|R_{\text{int}}^{(n)}\| &\leq \frac{1}{2} \left\|\sum_{i<j} [c_i^{(n)}H_i,c_j^{(n)}H_j] \right\| + \frac{1}{2N} \|[H,\Delta^{(n)}]\|\\
       &\leq \sum_{i<j} \|c_i^{(n)}H_i\| \|c_j^{(n)}H_j\| + \left\|\frac{H}{N}\right\| \|\Delta^{(n)}\|\\
       &\leq \frac{1}{2} \left[\left( \sum_{j=1}^M\|c_j^{(n)}H_j\| \right)^2 - \sum_{j=1}^M\|c_j^{(n)}H_j\|^2\right] + \delta^3\\
       &\leq \frac{\delta^2}{2}+\delta^3\\
       &\leq\delta^2,
    \end{split}
    \end{equation*}
    where we used
    $\|[A,B]\|\leq 2\|A\|\|B\|$ and $\delta\leq1/2$.
    This completes the induction.

    It remains to verify the hypotheses of Theorem~\ref{th: variable trotter error}.
    Equation~\eqref{eq: bound r int exact} already gives
    \begin{equation*}
        \sum_{j=1}^M\|c_j^{(n)}H_j\| \leq\delta,
    \end{equation*}
    while
    \begin{equation*}
        \left\|\frac{H}{N}\right\| \leq \sum_{j=1}^M\|\alpha_jH_j\| \leq \frac{\delta}{2}.
    \end{equation*}
    Moreover,
    \begin{equation*}
        \|\Delta^{(n)}\|\leq\delta^2.
    \end{equation*}
    Equation~\eqref{eq: local defect expansion} then gives
    \begin{equation*}
    \begin{split}
        \|X^{(n)}\| &\leq \|\Delta^{(n)}\| + \frac{1}{2} \left\|\sum_{i<j}[c_i^{(n)}H_i,c_j^{(n)}H_j]\right\| + \frac{1}{2N} \|[H,\Delta^{(n)}]\| + C\delta^3\\
        &\leq \delta^2+\frac{\delta^2}{2} + \delta^3+C\delta^3\\
        &\leq C\delta^2.
    \end{split}
    \end{equation*}
    Therefore,
    \begin{equation*}
        \sigma =  \sum_{n=1}^N\|X^{(n)}\| \leq CN\delta^2 \leq \delta_0.
    \end{equation*}
    Moreover,
    \begin{equation*}
        \sigma^3 \leq CN^2\delta^4,
    \end{equation*}
    where we used $CN\delta^2\leq\delta_0$ and absorbed $\delta_0$ into the universal constant.

    Finally, $R_{\text{int}}=R_{\text{int}}^{(N)}$, while
    \begin{equation*}
        \|R_{\text{ext}}\| \leq \frac{N(N-1)}{2} \max_{1\leq n\leq N}\|\Delta^{(n)}\|^2 \leq CN^2\delta^4.
    \end{equation*}
    Substituting these estimates into equations~\eqref{eq: full global error} and~\eqref{eq: remainder terms} gives
    \begin{equation*}
        \varepsilon \leq K_{\text{exact}} \left( \delta^2+N^2\delta^4+N\delta^3 \right),
    \end{equation*}
    completing the proof.
\end{proof}

We now move to the imperfect-synthesis case.

\begin{theorem}[Imperfect infinitesimal synthesis (finite $N$)]
    Suppose that $\calH$ is linearly independent and $\Lie(\calH)=\Span_{\R}\calH$.
    Let $\delta>0$ and $r>0$ satisfy
    \begin{equation}\label{eq: hypothesis alpha imperfect 1}
        2rN+C_0\delta^3N\leq \delta_0, \qquad 2\delta\leq\delta_0,
        \qquad\text{and}\qquad
        \sum_{j=1}^M\|\alpha_jH_j\|\leq \frac{\delta}{2},
    \end{equation}
    where $C_0$ and $\delta_0$ are universal constants.
    Assume moreover that the available Hamiltonians $\calH$ satisfy
    \begin{equation}\label{eq: hypothesis alpha imperfect 2}
        \kappa_{\calH} \leq \frac{\delta}{2\left(r+\eta(\calB(\alpha),r)\right)} \left(\sum_{j=1}^M\|H_j\|^2\right)^{-1/2}.
    \end{equation}
    Finally, take that the infimum in equation~\eqref{eq: eta} is attained for every $Y$ satisfying $\|Y-\calB(\alpha)\|\leq r$.

    If
    \begin{equation*}
        \eta(\calB(\alpha),r) + \frac{\delta^2}{2} + \delta\left(        r+\eta(\calB(\alpha),r) \right) \leq r,
    \end{equation*}
    then Algorithm~\ref{alg: schedule} produces a trotterization schedule
    whose total global error satisfies
    \begin{equation*}
        \varepsilon \leq K_{\text{imper}} \left(r+ N\delta^3 + N^2\left(\eta(\calB(\alpha),r)\right)^2 + N^2\delta^2 \left(r+\eta(\calB(\alpha),r)\right) + N^2\delta^4\right),
    \end{equation*}
    where $K_{\text{imper}}$ is a universal constant.
\end{theorem}

\begin{proof}
    The argument closely follows the proof of the exact-synthesis case.
    For convenience, set
    \begin{equation*}
        \eta_r:=\eta(\calB(\alpha),r), \qquad q:=r+\eta_r.
    \end{equation*}
    We first show inductively that $\|R_{\text{int}}^{(n)}\|\leq r$ for all $0\leq n\leq N$.
    The case $n=0$ is immediate, so assume $n\geq1$.

    Suppose that $\|R_{\text{int}}^{(n-1)}\|\leq r$.
    Since
    \begin{equation*}
        \left\|\calB(\alpha) - \Ad_{S^{n-1}}R_{\text{int}}^{(n-1)} - \calB(\alpha) \right\| \leq r,
    \end{equation*}
    the definition of $\eta_r$, together with the attainment assumption, guarantees an admissible $c^{(n)}\in\calC$ such that
    \begin{equation*}
        \left\| \Delta^{(n)} + \Ad_{S^{n-1}}R_{\text{int}}^{(n-1)} \right\| \leq\eta_r.
    \end{equation*}
    Consequently,
    \begin{equation}\label{eq: imperfect drift bound}
        \|\Delta^{(n)}\| \leq r+\eta_r=q.
    \end{equation}

    Since $\calH$ is linearly independent, the synthesis map $\calB$ is injective.
    Therefore, $c^{(n)}-\alpha$ is the unique coefficient vector satisfying
    \begin{equation*}
        \calB(c^{(n)}-\alpha)=\Delta^{(n)},
    \end{equation*}
    and the definition of $\kappa_{\calH}$ gives
    \begin{equation*}
        \|c^{(n)}-\alpha\| \leq  \kappa_{\calH}\|\Delta^{(n)}\| \leq \kappa_{\calH}q.
    \end{equation*}
    Hence, by equation~\eqref{eq: hypothesis alpha imperfect 2},
    \begin{equation*}
    \begin{split}
        \sum_{j=1}^M\|c_j^{(n)}H_j\| &\leq \sum_{j=1}^M\|\alpha_jH_j\| + \sum_{j=1}^M |c_j^{(n)}-\alpha_j|\|H_j\|\\ 
        &\leq \frac{\delta}{2} + \|c^{(n)}-\alpha\| \sqrt{\sum_{j=1}^M\|H_j\|^2}\\
        &\leq \frac{\delta}{2} + \kappa_{\calH}q \sqrt{\sum_{j=1}^M\|H_j\|^2}\\
        &\leq\delta.
    \end{split}
    \end{equation*}
    Hence,
    \begin{equation*}
        \left\|
        \frac{1}{2}\sum_{i<j}
        [c_i^{(n)}H_i,c_j^{(n)}H_j]
        \right\|
        \leq\frac{\delta^2}{2}.
    \end{equation*}

    Using equation~\eqref{eq: r int} and unitary invariance,
    \begin{equation*}
    \begin{split}
        \|R_{\text{int}}^{(n)}\|
        &\leq \eta_r +
        \frac{\delta^2}{2}  + \delta q\\
        &\leq r,
    \end{split}
    \end{equation*}
    where the last inequality is precisely equation~\eqref{eq: imperfect invariant radius}.
    This completes the induction.

    We now verify the remaining hypotheses of Theorem~\ref{th: variable trotter error}.
    By equations~\eqref{eq: local defect expansion} and~\eqref{eq: imperfect drift bound},
    \begin{equation*}
    \begin{split}
        \|X^{(n)}\| &\leq q + \frac{\delta^2}{2} + \delta q + C\delta^3\\
        &\leq 2r+C\delta^3,
    \end{split}
    \end{equation*}
    where we again used equation~\eqref{eq: imperfect invariant radius}.
    Therefore,
    \begin{equation*}
        \sigma =  \sum_{n=1}^N\|X^{(n)}\| \leq  2rN+C\delta^3N \leq\delta_0,
    \end{equation*}
    after choosing $C_0$ in equation~\eqref{eq: hypothesis alpha imperfect 1} sufficiently large.

    Furthermore,
    \begin{equation*}
        \max_{1\leq n\leq N}\|\Delta^{(n)}\| \leq q,
    \end{equation*}
    and hence
    \begin{equation*}
        \|R_{\text{ext}}\| \leq CN^2q^2.
    \end{equation*}
    The remainder estimate of Theorem~\ref{th: variable trotter error} gives
    \begin{equation*}
        \|R_3\| \leq C\left( N\delta^3 +  N^2\delta^2q + N^2\delta^4 + \sigma^3 \right).
    \end{equation*}

    Since equation~\eqref{eq: imperfect invariant radius} implies $\eta_r\leq r$, we have $q\leq2r$.
    Together with $2rN+C_0\delta^3N\leq\delta_0$, this allows $\sigma^3$ to be absorbed into
    \begin{equation*}
        C\left(N^2q^2+N\delta^3\right).
    \end{equation*}
    Finally, applying Theorem~\ref{th: variable trotter error} and using
    $\|R_{\text{int}}^{(N)}\|\leq r$ gives
    \begin{equation*}
        \varepsilon \leq K_{\text{imper}} \left(r + N\delta^3 + N^2q^2 + N^2\delta^2q + N^2\delta^4 \right),
    \end{equation*}
    completing the proof.
\end{proof}

\section{Proof of Proposition~\ref{prop: expansion}}\label{app: proof expansion}
\begin{proof}[Proof of Proposition~\ref{prop: expansion}]
    We first note that conjugation preserves $\su(d)$. In particular, for every
    $U\in\SU(d)$ and $X\in\su(d)$,
    \begin{equation*}
        \exp\left(i\Ad_U X\right)  =  U\exp(iX)U^{-1},
    \end{equation*}
    and in particular the operator $H_{\texttt{new}} = \Ad_{\exp(i\tau H_i)}H_j$ constructed in Algorithm~\ref{alg: expand} belongs to $\su(d)$.

    It remains to show that, whenever
    \begin{equation*}
        [H_i,H_j]\notin\Span_{\R}\calH,
    \end{equation*}
    there exists a sufficiently small $\tau>0$ such that $H_{\texttt{new}}\notin\Span_{\R}\calH$.

    Let $\Pi$ denote the orthogonal projection onto $\Span_{\R}\calH$ and define
    \begin{equation*}
        F(t):=(I-\Pi)\Ad_{\exp(iH_it)}H_j.
    \end{equation*}
    Since $H_j\in\Span_{\R}\calH$, we have $F(0)=0$. On the other hand,
    \begin{equation*}
        F'(0) = (I-\Pi)[H_i,H_j] \neq 0,
    \end{equation*}
    where we used the definition of the adjoint representation and the assumption $[H_i,H_j]\notin\Span_{\R}\calH$.
    Therefore, by differentiability, $F(t)\neq0$ for all sufficiently small nonzero $t$.
    Hence, $ \Ad_{\exp(iH_it)}H_j\notin\Span_{\R}\calH$ for all sufficiently small nonzero $t$.
    Thus the sequence $\tau=2^{-k}$ eventually produces a Hamiltonian that increases the dimension of $\Span_{\R}\calH$ by one.
    Every such Hamiltonian remains in $\Lie(\Span_{\R}\calH)$, since the Lie  closure is invariant under the adjoint action generated by its elements.
    Consequently, each successful augmentation increases $\dim\Span_{\R}\calH$ by one without leaving its Lie closure.

    The procedure therefore terminates when $\Span_{\R}\calH=\Lie(\Span_{\R}\calH),$ after exactly $$
        \calM = \dim\Lie(\Span_{\R}\calH)-\dim\Span_{\R}\calH $$
    successful augmentations.
\end{proof}

\end{document}